\documentclass[twocolumn,showpacs,showkeys,preprintnumbers,amsmath,amssymb,pra]{revtex4}
\usepackage{graphicx}% Include figure files
\usepackage{dcolumn}% Align table columns on decimal point
\usepackage{bm}% bold math

\newcommand{\mcal}{\ensuremath{\mathcal}}
\newcommand{\opr}[1]{\ensuremath{\mathbf{\mathsf{#1}}}}

\newcommand{\fprod}[2]{\ensuremath{\left.\left<#1\right|#2\right>}}

\newcommand{\expo}[1]{\ensuremath{\mbox{e}^{#1}}}
\newcommand{\abs}[1]{\ensuremath{\left|#1\right|}}

\newcommand{\ket}[1]{\ensuremath{\left|\left. #1 \right>\right.}}

\begin{document}

%\preprint{Final Draft-01}

\title{Confined Quantum Time of Arrival for Vanishing Potential}% Force line breaks with \\

\author{Eric A. Galapon$^{1,2,3}$}
\email{eric.galapon@up.edu.ph}
\author{Roland F. Caballar$^{1}$}
%\email{rcaballarr@nip.upd.edu.ph}
\author{Ricardo Bahague$^1$}
%\email{rbahague@nip.upd.edu.ph}
\affiliation{$^1$Theoretical Physics Group, National Institute of Physics, University of
 the Philippines, Diliman, Quezon City, 1101 Philippines}
\affiliation{$^2$Theoretical Physics, The University of the Basque Country, Apdo. 
644, 48080 Bilbao, Spain}
\affiliation{$^3$Chemical Physics, The University of the Basque Country, Apdo. 
644, 48080 Bilbao, Spain}

\date{\today}% It is always \today, today,

\begin{abstract}
We give full account of our recent report in [E.A. Galapon, R. Caballar, R. Bahague {\it Phys. Rev. Let.} {\bf 93} 180406 (2004)] where it is shown that formulating the free quantum time of arrival problem in a segment of the real line suggests rephrasing the quantum time of arrival problem to finding a complete set of states that evolve to unitarily arrive at a given point at a definite time. For a spatially confined particle, here it is shown explicitly that the problem admits a solution in the form of an eigenvalue problem of a class of compact and self-adjoint time of arrival operators derived by a quantization of the classical time of arrival. The eigenfunctions of these operators are numerically demonstrated to unitarilly arrive at the origin at their respective eigenvalues.
\end{abstract}

%Use showkeys class option if keyword
\maketitle

\section{Introduction}

The quantum free time of arrival problem has been the subject of numerous investigations in the past and in current times \cite{pauli,bhor,paul,allc,raza,kijo,jammer,olhovsky,park,grot,blanchard,self,sri,busch1,busch2,busch4,povm,egus,toller,busch3,muga,leo}. The past has been mostly marked by stark pessimism fueled by Pauli's well-known theorem on the non-existence of self-adjoint time operators \cite{pauli}, and Allcock's resigned conclusion that no ideal time of arrival distribution exists within the standard framework of quantum mechanics \cite{allc}. The current times, in contrast, is marked by optimism fueled by the generalization of quantum observables to include positive operator valued measures (POVM) \cite{sri,busch1,busch2,busch4,povm,egus,toller,busch3,gian,gen}, and the advent of Kijowski's ideal quantum time of arrival distribution \cite{kijo}. This optimism has been further strengthened by the realization that Kijowski's distribution can be completely derived from a POVM arising from the quantization of the classical free time of arrival \cite{egus}. And this optimism has been capped by the recognition that Kijowski's distribution has an operational meaning \cite{dambo,baut,heg,baut2,heg1}. These recent significant developments have had led to the impression that the quantum free time of arrival problem has been solved, and only its proper generalization and experimental realization are left undone \cite{private}.

While the above developments have unfolded, an unexpected development has been in the silent offing. First, it was demonstrated by one of us that Pauli's theorem does not hold within the single Hilbert space formulation of quantum mechanics, and showed the consistency of self-adjoint and bounded time operators canonically conjugate with a semibounded discrete Hamiltonian \cite{galapon,galapon2,galapon3}. This in turn has led to the realization that the non-self-adjointness of the quantized free time of arrival (TOA) operator has nothing to do with the semiboundedness of the Hamiltonian \cite{galapon}, as Pauli's theorem would assert otherwise. This then led to the unexpected result that the non-self-adjointness of the TOA-operator can in fact be lifted by spatial confinement. Thus the concept of confined quantum time of arrival (CTOA) was introduced \cite{galapon4}. The CTOA-operators form a class of compact and self-adjoint operators canonically conjugate with their respective Hamiltonians in a non-dense subspace of the system Hilbert space. Being compact, the CTOA-operators posses discrete spectrum, and a complete set of mutually orthogonal square integrable eigenfunctions. However, the interpretation of the spectral properties of the CTOA-operators was not clear. Thus in \cite{galapon0} we addressed the issue of interpretation and it entailed rephrasing the quantum time of arrival problem in finding a complete set of states that unitarily arrive at some predetermined point. The eigenfunctions of the CTOA-operators are found to be states that evolve to unitarily arrive at the origin at their respective eigenvalues. This result has repercussions beyond the quantum time of arrival problem which we will address elsewhere.

In this paper we give full account of the confined quantum time of arrivals for vanishing potentials. In Section-{\ref{toareal}} we give a short review of the time of arrival operator in the real line. In Section-\ref{sctoa} we give the detailed formulation of the confined quantum time of arrival operators. In Section-\ref{conj} we investigate the structure of the conjugacy of the CTOA-operators with their Hamiltonians. In Section-\ref{sym} we study the symmetries of the CTOA operators and from them find the general properties of their eigenfunctions and eigenvalues. In Section-\ref{eigprob} we solve analytically the CTOA-operator eigenvalue problem. In Section-\ref{dynamics} we study numerically the dynamical properties of the eigenfunctions and provide unambiguous interpretation for the spectral properties of the CTOA-operators. In Section-\ref{conclu} we discuss the physical repercussions of our current results.

\section{Time of Arrival Operator in the real line}\label{toareal}
The QTOA-problem is traditionally the problem of finding the time of arrival  (TOA) distribution of a structureless particle prepared in some initial state at a given point, say, at the origin. This operator is presumed to be the quantized classical-TOA in unbounded free space. That is, if a classical free particle, of mass $\mu$ in one dimension at location $q$ with momentum $p$, will arrive at the origin at the time $T(q,p)=-\mu qp^{-1}$, then the quantum TOA-distribution must be derivable from the quantization of $T(q,p)$, from the operator 
\begin{equation}
\opr{T}=-\frac{1}{2}\mu (\opr{qp^{-1}+p^{-1}q}).\label{qtoa}
\end{equation}
Formally the time of arrival operator $\opr{T}$ is canonically conjugate to the free Hamiltonian, $\opr{H}=(2\mu)^{-1}\opr{p}^2$, i.e. $[\opr{H},\opr{T}]=i\hbar$. 

Much of the study on equation-(\ref{qtoa}) has been done in momentum representation, in which it assumes the formal form
\begin{equation}
\opr{T}=\frac{i\hbar \mu}2\left(\frac1{p^2}-\frac2p
\frac{\partial}{\partial p}\right)\,.\nonumber
\end{equation}
As disccused in \cite{egus}, $\opr{T}$ is a densely defined, unbounded operator in ${\cal
H}_{p}:=L^{2}\left({\bf R},{\rm d}p\right)$. And it has the following degenerate non-square integrable eigenfunctions
\[\tilde\psi^{(t)}_{\alpha}(p)=\Theta(\alpha p)\left(\frac{\alpha
p}{2\pi \mu \hbar}\right)^{1/2} e^{i p^2 t/2 \mu\hbar}\,\]
where $\alpha=\pm 1$. This set of eigenfunctions is complete, i.e.,$\sum_{\alpha}\int_{-\infty}^\infty{\rm
d}t\,\tilde\psi_{\alpha}^{(t)}(p')\tilde\psi_{\alpha}^{(t)}(p)=\delta(p-p')$. However, they are nonorthogonal, i.e. \[\int_{-\infty}^\infty{\rm d}p\,\tilde\psi^{(t')}_{\alpha'}(p)\tilde\psi^{(t)}_{\alpha}(p)=\frac12
\delta_{\alpha\alpha'}\left(\delta(t-t')+\frac{i}{\pi}{\rm P}\frac{1}{t-t'}\right)\,.\]
This nonorthogonality is a reflection of the non-self-adjointness of $\opr{T}$ in the real line. In fact $\opr{T}$ is maximally symmetric so that it has no self-adjoint extension. The space-time properties of the TOA-operator eigenfunctions are studied in \cite{toa}.

For a long time the non-self-adjointness of equation-(\ref{qtoa}) in the real line has been construed as a consequence of Pauli's theorem, in particular, from the semiboundedness of the Hamiltonian. However, as we will show in the following section, the non-self-adjointness has nothing to do with the semiboundedness of the Hamiltonian.

\section{The Confined Time of Arrival Operators}\label{sctoa}
\subsection{The System}
 Let the particle be confined between two points with length $2l$. If $p\neq 0$ and $\abs{q}<l$, the classical time of arrival at the origin (the first time of arrival, i.e. arrival without reflection from the boundaries) and the Hamiltonian are still  given by $T=-\mu qp^{-1}$ and $H=(2\mu)^{-1}p^2$, respectively; moreover, $T$ remains canonically conjugate with the  Hamiltonian. Then equation-(\ref{qtoa}) is still the totally symmetric  quantized form of $T$ even when the particle is confined, and it likewise  remains canonically conjugate with the Hamiltonian.

To give meaning to $\opr{T}$ for the spatially confined particle, we attach the Hilbert space $\mcal{H}=L^2[-l,l]$ to the system. The position  operator is unique and is  given by the bounded operator $\opr{q}$,  $\left(\opr{q}\varphi\right)\!(q)=q\varphi(q)$ for all $\varphi(q)$ in $\mcal{H}$. On the other hand, the momentum operator and the Hamiltonian are not unique, and have to be considered carefully.  Our choice is dictated by the assumption of closedness of the system and the requirement of consistency with quantization: We assume the system to be conservative and we require that the evolution of the system be generated by a purely kinetic Hamiltonian. The former requires a self-adjoint Hamiltonian to ensure that time evolution is unitary. The later requires a self-adjoint momentum operator commuting with the Hamiltonian to ensure that the quantum Hamiltonian is the quantization of the purely kinetic Hamiltonian of the freely evolving classical particle between the boundaries. 

One of the possible Hamiltonians that can be assigned to the system is the textbook Hamiltonian where the domain of the Hamiltonian operator is restricted to those vectors that vanish at the boundaries. But this Hamiltonian, while self-adjoint, does not satisfy the second requirement: No self-adjoint momentum operator commuting with the Hamiltonian exists. The reason for this is that the eigenfunctions of self-adjoint momentum operator in a bounded segment of the line are plane waves and none of these eigenfunctions vanishes at the boundaries (see below). This means that the momentum operator and the Hamiltonian do not have a common set of eigenvectors. The Hamiltonian then can not be purely kinetic. For this reason, we abandon this Hamiltonian and consider another. 

Now for every $\gamma$ in the interval $(-\pi/2,\leq\pi/2]$, there exists a self-adjoint momentum operator given by  the operator $\opr{p_{\gamma}}=-i\hbar \partial_{q}$ whose domain consists of those vectors $\phi(q)$ in $\mcal{H}$ with square integrable first derivatives, i.e. $\int\abs{\phi'(q)}^2\,dq<\infty$, satisfying the boundary condition  $\phi(-l)=\expo{-2i\gamma}\phi(l)$. With $\opr{p}_{\gamma}$ self-adjoint, the kinetic energy operator $\opr{K_{\gamma}}=\frac{1}{2\mu}\opr{p}_{\gamma}^2$ is consequently self-adjoint. Thus the Hamiltonian is purely kinetic,
\begin{equation}
\opr{H}_{\gamma}=\frac{1}{2\mu}\opr{p}_{\gamma}^2,
\end{equation}
the domain of which consists of all vectors $\phi(q)$ in the domain of the momentum operator $\opr{p}_{\gamma}$ such $(\opr{p}_{\gamma}\phi)\!(q)$ is still in the domain of $\opr{p}_{\gamma}$. The momentum and the Hamiltonian then commute and have the common set of plane wave eigenvectors \begin{equation}
\phi_k^{(\gamma)}(q)=\frac{1}{\sqrt{2l}} \exp\!\!\left[i\,(\gamma+k \pi)\frac{q}{l}\right],
\end{equation}
where $k=0,\pm1,\pm2\cdots$; and their respective eigenvalues are $p_{\gamma,k}=\hbar (\gamma+k\pi)l^{-1}$ and $E_{\gamma,k}=\hbar^2(\gamma+k\pi)^2(2\mu l^2)^{-1}$.

One may ask which of these infinitely many Hamiltonians should we use in constructing the operators corresponding to the confined classical time of arrival. We will find below that we have to consider the whole window $(-\pi,\pi/2]$ in order to cover the entire symmetry of the classical time of arrival enumerated above in the quantum domain. Likewise, we shall show that for every $\gamma$ Equation-(\ref{qtoa}) defines a self-adjoint operator, $\opr{T_{\gamma}}$, satisfying the canonical commutation relation with the Hamiltonian $\opr{H_{\gamma}}$ in a closed subspace of the Hilbert space. And that the operators $\opr{T_{\gamma}}$ can be legitimately called as time of arrival operators; this follows from our study of the dynamics of the eigenfunctions of the TOA-operators. 

\subsection{Non-periodic Confined Time of Arrival Operators}
Now let us consider $\opr{T}$ for $\gamma\neq 0$. Since  $\opr{q}$ appears in first power in $\opr{T}$, $\opr{T}$ is an operator if  the inverse of $\opr{p}_{\gamma}$ exists.  Since zero is not an eigenvalue of  $\opr{p}_{\gamma}$, the inverse $\opr{p_{\gamma}^{-1}}$ exists, and is in fact bounded and  self-adjoint. Then it follows that, for every $\gamma$, $\opr{T}$ is a  bounded, symmetric operator. Thus $\opr{T}$ is self-adjoint. For a given $\gamma$, we identify $\opr{T}$ with the operator
\begin{equation}
\opr{T}_{\gamma}=-\frac{1}{2}\mu(\opr{q}\opr{p_{\gamma}^{-1}}+\opr{p_{\gamma}^{-1}}\opr{q}),
\end{equation}
derived from the formal operator $\opr{T}$ by replacing $\opr{p}$ with $\opr{p}_{\gamma}$. We shall refer to $\opr{T}_{\gamma}$ as the non-periodic confined time of arrival (CTOA) operator for a given $\abs{\gamma}<\pi$. 

In coordinate representation, $\opr{T}_{\gamma}$ assumes the form of a Fredholm integral operator 
\begin{equation}
(T_{\gamma}\varphi)(q)=\int_{-l}^{l}\left<q\right|\opr{T}_{\gamma}\left|q'\right>\varphi(q') dq'.\label{nonperiodiccase}
\end{equation}
where the kernel is given by
\begin{equation} \label{repre}
\left<q\right|\opr{T}_{\gamma}\left|q'\right>=-\mu\frac{(q+q')}{4\hbar\sin\gamma}\left(e^{i\gamma}
\mbox{H}(q-q')+e^{-i\gamma}\mbox{H}(q'-q) \right),
\end{equation}
in which H$(q-q')$ is the Heaviside function (see Appendix-A for the derivation of the kernel). The kernel $\left<q\right|\opr{T}_{\gamma}\left|q'\right>$ is both symmetric and square integrable, i.e. $\left<q\right|\opr{T}_{\gamma}\left|q'\right>=\left<q'\right|\opr{T}_{\gamma}\left|q\right>^{*}$ and $\int\int \abs{\left<q\right|\opr{T}_{\gamma}\left|q'\right>}^2\,dq\,dq'<\infty$, respectively. This reaffirms the self-adjointness of $\opr{T_{\gamma}}$.  These two properties imply that $\opr{T_{\gamma}}$ is a compact, self-adjoint operator: It possesses a complete set of eigenfunctions with a corresponding discrete set of eigenvalues. 

\subsection{Periodic Confined Time of Arrival Operator}

For the periodic case $\gamma=0$, $\opr{p^{-1}}$ is ill-defined because $\opr{p}$ has no inverse, the zero being an eigenvalue of $\opr{p}$. But this can be remedied. The pathology arises from the one-dimensional subspace spanned by the state of vanishing momentum, the null subspace $\mcal{N}(\opr{p})$. But this subspace has no bearing to the problem because the question when a given particle arrives makes sense only when the particle is in motion. We expect then that $\opr{T}$ is well defined if the contribution of the null subspace is removed. 
 
Technically this can be accomplished as follows \cite{galapon}: Let $\opr{E}$ and $\opr{E}^{\perp}$ be the projections unto the closures of the subspaces $\mathcal{N}(\opr{p})$ (the subspace spanned by the zero momentum state) and $\mcal{N}(\opr{p})^{\perp}$ (the subspace spanned by the non-vanishing momentum states), respectively. Now $\mcal{N}(\opr{p})$ and $\mcal{N}(\opr{p})^{\perp}$ are invariant under $\opr{p}$; both subspaces then reduce  $\opr{p}$. Because $\opr{p}$ is self-adjoint, its restrictions on $\mcal{N}(p)$ and $\mcal{N}(p)^{\perp}$  are both self-adjoint. The restriction $\opr{p}_{\mcal{N}^{\perp}}$ has trivial null-space, so that its inverse, $\opr{p}_{\mcal{N}^{\perp}}^{-1}$, exists in $\mcal{P}^{\perp}\mcal{H}$. But this inverse exists only in $\mcal{P}^{\perp}\mcal{H}$ and not in $\mcal{H}$. This can be addressed by extending 
$\opr{p}_{\mcal{N}^{\perp}}^{-1}$ in the entire $\mcal{H}$. First we note that $\opr{p}_ {\mcal{N}^{\perp}}^{-1}$ is self-adjoint and bounded. It can be shown that  $\opr{p}_{\mcal{N}^{\perp}}^{-1}$ is subnormal and it admits a unique minimal  extension in the entire $\mcal{H}$. Its minimal extension is the bounded and self-adjoint operator $\opr{P}^{-1}=\opr{E}^{\perp} \opr{p}_{\mcal{N}^{\perp}}^{-1}\opr{E}^{\perp}$. This operator can be interpreted as the quantization of the classical observable $p^{-1}$ for $p\neq 0$ under the boundary condition imposed upon the momentum operator.

Substituting $\opr{P^{-1}}$ for $\opr{p^{-1}}$ in the formal time of arrival operator, we get the operator quantization of $T$ for periodic boundary condition,
\begin{equation}
\opr{T}_0=-\frac{1}{2}\mu(\opr{q}\opr{P}^{-1}+\opr{P}^{-1}\opr{q}).
\end{equation} 
Note that both $\opr{q}$ and $\opr{P^{-1}}$ are bounded, everywhere defined, self-adjoint operators. Since $\opr{T}_0$ is symmetric under the exchange of $\opr{q}$ and $\opr{P^{-1}}$, it is likewise bounded, everywhere defined and self-adjoint. We shall refer to $\opr{T}_0$ as the periodic confined quantum time of arrival operator. In position representation, $\opr{T}_0$ likewise assumes the form of a Fredholm integral operator,
\begin{equation}
(T_{0}\varphi)\!(q)=\int_{-l}^{l}\left<q\right|\opr{T}_{0}\left|q'\right>\varphi(q') dq'.\label{periodiccase}
\end{equation}
whose kernel is
\begin{equation} \label{periodic}
	\left<q\right|\opr{T}_0\left|q'\right>=\frac{\mu}{4i \, \hbar}(q+q')\mbox{sgn}(q-q')-\frac{\mu}{4i\,\hbar
 l}\left(q^2 -q'^2\right).
\end{equation}
The kernel $\left<q\right|\opr{T}_0\left|q'\right>$ is likewise symmetric and square integrable. This means that the finite periodic limit of (\ref{repre}) generates a self-adjoint integral operator, $\opr{T_0}$, whose kernel is given by equation (\ref{periodic}).  This operator is likewise compact---and thus discrete. And its eigenfunctions form a complete set of orthonormal system.

\subsection{The Noncovariance of the CTOA-operators}
A time operator $\opr{T}$ is covariant if its eigenvectors, $\left|\left. \tau\right>\right.$, satisfy the property $e^{-i\opr{H}t/\hbar}\left|\left. \tau\right>\right.=\left|\left. \tau-t\right>\right.$, where $\left|\left. \tau-t\right>\right.$ is also an eigenvector of $\opr{T}$ for any time $t$. Covariance of $\opr{T}$ implies that it has a completely continuous spectrum taking values in the entire real line. The time of arrival operator in $L^2(-\infty,\infty)$, for example, is covariant. Since the confined time of arrival operators posses a pure point spectrum, they are not covariant. Covariance has been a premium requirement imposed upon time operators \cite{busch1,busch2,busch3,povm,egus,toller,gian,atm}. We will, however, demonstrate in the following sections that non-covariant time operators are physically meaningful.

\section{The Conjugacy of the CTOA-operators with their Hamiltonians}\label{conj}

\subsection{Non-periodic case}
The commutator between $\opr{T}_{\gamma}$ and $\opr{H}_{\gamma}$ in the system Hilbert space is defined only if there exists a non-trivial intersection between the domains of the the composition operators $\opr{T}_{\gamma}\opr{H}_{\gamma}$ and $\opr{H}_{\gamma}\opr{T}_{\gamma}$. It is not necessary that the commutator domain---the subspace $\mcal{D}_{com}$ in which the operator $(\opr{T}_{\gamma}\opr{H}_{\gamma}-\opr{H}_{\gamma}\opr{T}_{\gamma})$ is defined---coincides with the canonical domain---the subspace $\mcal{D}_{can}$ in which the operator $(\opr{T}_{\gamma}\opr{H}_{\gamma}-\opr{H}_{\gamma}\opr{T}_{\gamma})$ is proportional to the identity operator in $\mcal{D}_{can}$. Generally we have the inclusion relation $\mcal{D}_{can}\subseteq\mcal{D}_{com}$ for any pair of operators. In our case, we will find that $\mcal{D}_{can}$ is a proper subspace of $\mcal{D}_{com}$.

First let us find the domain of $\opr{H}_{\gamma}\opr{T}_{\gamma}$. Since $\opr{T}_{\gamma}$ is bounded, and thus defined in the entire Hilbert space, the domain of $\opr{H}_{\gamma}\opr{T}_{\gamma}$ consists of all vectors $\phi$ in the Hilbert such that $\opr{T}_{\gamma}\phi$ is in the domain of the Hamiltonian. Recall that the domain of $\opr{H}_{\gamma}$ consists of those that satisfy some boundary conditions (see below). Let $\varphi(q)=\left(\opr{T}_{\gamma}\phi\right)\!(q)$, i.e.
\begin{eqnarray}
\varphi(q)&=&-\frac{\mu}{4\hbar \sin\gamma}e^{i\gamma}\int_{-l}^q (q+q')\phi(q') dq'\nonumber\\
 & &-\frac{\mu}{4\hbar \sin\gamma}e^{-i\gamma}\int_q^l (q+q')\phi(q') dq' \label{koo}
\end{eqnarray}

If $\varphi(q)$ were to be in the domain of the Hamiltonian, first it must satisfy the boundary condition $\varphi(-l)=e^{-2 i \gamma}\varphi(l)$. Evaluating equation-(\ref{koo}) at the boundaries yield
\begin{eqnarray}
\varphi(l)&=&-\frac{\mu e^{i\gamma}}{4\hbar \sin\gamma}\int_{-l}^l (l+q')\phi(q') dq'\nonumber\\
\varphi(-l)&=&-\frac{\mu e^{-i\gamma}}{4\hbar \sin\gamma}\int_{-l}^l (-l+q')\phi(q') dq'\nonumber
\end{eqnarray}
Imposing the first boundary condition on $\varphi$, gives us the equality $\int_{-l}^l\phi(q') dq'=-\int_{-l}^l \phi(q') dq'$, which is only true if and only if both sides are equal to zero. Then $\phi$ must satisfy the condition $\int_{-l}^l\phi(q') dq'=0$. That is the domain of $\opr{H}_{\gamma}\opr{T}_{\gamma}$ is orthogonal to the subspace spanned by the zero-momentum-eigenfunction.

Moreover, $\varphi(q)=\left(\opr{T}_{\gamma}\phi\right)\!(q)$ must satisfy the second boundary condition $\varphi'(-l)=e^{-2 i \gamma}\varphi'(l)$. Taking the first derivative of equation-(\ref{koo}) gives
\begin{eqnarray}
\varphi'(q)\!&=&\!-\frac{\mu}{4\hbar \sin\gamma}\left[e^{i\gamma}\int_{-l}^q \phi(q') dq'
 + e^{-i\gamma}\int_q^l \phi(q') dq' \right]\nonumber\\
 & &\hspace{3cm}-i\frac{\mu}{\hbar} \,q\,\phi(q).\label{diff1}
\end{eqnarray}
Because $\phi(q)$ must satisfy $\int_{-l}^l\phi(q') dq'=0$, the values at the boundaries of the derivative simplifies to $\varphi'(l)=-i\mu\hbar^{-1}l \phi(l)$ and $\varphi'(-l)=i\mu\hbar^{-1}l \phi(-l)$.
Imposing the second boundary condition gives us $\phi(-l)=-e^{-2i \gamma}\phi(l)$. Note that there infinitely many vectors satisfying these in the domain of the Hamiltonian, and vectors that lie outside the domain of $\opr{H}_{\gamma}$.

The domain of the operator $\opr{H}_{\gamma}\opr{T}_{\gamma}$ then consists of all vectors $\phi(q)$ in the Hilbert space satisfying the conditions $\int_{-l}^l\phi(q') dq'=0$ and $\phi(-l)=-e^{-2i \gamma}\phi(l)$. Because of the first condition, the domain is orthogonal to the one dimensional subspace spanned by the zero momentum eigenvector. The operator $\opr{H}_{\gamma}\opr{T}_{\gamma}$ is then not densely defined. For all vectors $\phi$ in this domain, $\opr{H}_{\gamma}\opr{T}_{\gamma}$ acts as
\begin{equation}
\left(\opr{H}_{\gamma}\opr{T}_{\gamma}\phi\right)\!(q)=\frac{3}{4}i\hbar\phi(q)+\frac{1}{2}i\hbar q \phi'(q).
\end{equation}
We arrive at this expression by further differentiating Equation-(\ref{diff1}) and by multiplying the appropriate constants.

On the other hand, the domain of $\opr{T}_{\gamma}\opr{H}_{\gamma}$ consists of all vectors $\varphi$ in the domain of $\opr{H}_{\gamma}$ such that $\opr{H}_{\gamma}\varphi$ is in the domain of $\opr{T}_{\gamma}$. But since $\opr{T}_{\gamma}$ is bounded, taking the entire Hilbert space as its domain, the vector $\opr{H}_{\gamma}\varphi$ is automatically in the domain of $\opr{T}_{\gamma}$. The domain of $\opr{T}_{\gamma}\opr{H}_{\gamma}$ is then the entire domain of the Hamiltonian. In this case $\opr{T}_{\gamma}\opr{H}_{\gamma}$ is densely defined because the domain of the Hamiltonian is dense. In this domain, the operator $\opr{T}_{\gamma}\opr{H}_{\gamma}$ acts as
\begin{eqnarray}
\left(\opr{T}_{\gamma}\opr{H}_{\gamma}\phi\right)\!(q)\!=\!-\frac{1}{4}i \hbar \phi(q) \!+\! \frac{1}{2}i\hbar q \phi'(q)\!+\! \frac{1}{4}\hbar l e^{i\gamma} \phi'(-l),
\end{eqnarray}
where two successive integration by parts have been made to arrive at this expression, and the boundary condition on the elements of the domain of $\opr{H}_{\gamma}$ has been imposed in the simplification.

Now the commutator $\left[\opr{T}_{\gamma},\opr{H}_{\gamma}\right]=\opr{T}_{\gamma}\opr{H}_{\gamma}-\opr{H}_{\gamma}\opr{T}_{\gamma}$ is defined only on the subspace of the Hilbert space which is the intersection of the domains of the operators $\opr{H}_{\gamma}\opr{T}_{\gamma}$ and $\opr{T}_{\gamma}\opr{H}_{\gamma}$. The vectors in the domain of $\opr{H}_{\gamma}\opr{T}_{\gamma}$ satisfy the boundary condition $\phi(-l)=-e^{-2i \gamma}\phi(l)$; while those in the domain of $\opr{T}_{\gamma}\opr{H}_{\gamma}$ satisfy $\phi(-l)=e^{-2i \gamma}\phi(l)$. In order for these two boundary conditions to be satisfied simultaneously, we must have $\phi(-l)=\phi(l)=0$. Then the commutator domain consists of all vectors, $\phi$ in the domain of the Hamiltonian satisfying the conditions $\int_{-l}^l\phi(q') dq'=0$ and  $ \phi(l)=\phi(-l)=0$. In this domain,  the commutator of $\opr{H}_{\gamma}$ and $\opr{T}_{\gamma}$ is 
\begin{equation}
\left((\opr{H_{\gamma}T_{\gamma}}-\opr{T_{\gamma}H_{\gamma}})\phi\right)\!\!(q)=i\,\hbar\,\phi(q)
	+ \frac{1}{2}\hbar l e^{i\gamma} \phi'(-l).
\end{equation}

Clearly $\opr{H}_{\gamma}$ and $\opr{T}_{\gamma}$ are not canonically conjugate in the entire commutator domain. However, restricting the domain to those whose first derivatives vanish at the boundaries gives us a canonical domain. Thus in the subspace of the domain of the Hamiltonian consisting of all vectors $\phi(q)$ satisfying
\begin{equation}
\int_{-l}^l\!\!\phi(q') dq'=0,\;\;\phi^{(k)}(\pm l)=0 \;\;\mbox{for}\;\; k=0,1,
\end{equation}
the Hamiltonian and the confined time of arrival operator are canonically conjugate,
\begin{equation}
\left((\opr{H_{\gamma}T_{\gamma}}-\opr{T_{\gamma}H_{\gamma}})\phi\right)\!\!(q)=i\,\hbar\,\phi(q),
\end{equation}
and they are conjugate in a non-dense subspace, which is not the usual for canonical pairs.

\subsection{Periodic case}
Following the same steps above, we find that $\opr{H}_0$ and $\opr{T}_0$ form a canonical pair in a non-dense subspace of the Hilbert space consisting of the vectors satisfying the conditions
\begin{equation}
		\varphi^{(k)}(\pm l)=0,\;\;\;
				\int_{-l}^{l} q^{k}\,\varphi(q)\,dq=0 \;\;\mbox{for}\;\; k=0,1.
\end{equation}
That is
\begin{equation}
		\left(\left(\opr{H_0 T_0}-\opr{T_0 H_0}\right)\varphi\right)\!\!(q)=i\hbar\,\varphi(q)
\end{equation}
for all $\varphi$ in the canonical domain. Being orthogonal to the two-dimensional subspace whose elements are  $\varphi(q)=a+bq$ for complex $a$ and $b$, the canonical domain is not dense. As in the former case, the canonical domain is smaller than the commutator domain of $\opr{H}_0$ and $\opr{T}_0$.

\subsection{Quantum canonical pairs}

The above prescribed quantization of the Hamiltonian and the time of arrival for the spatially confined particle yields the correspondence 
\begin{eqnarray}
\left\{T,H\right\}=1\;\;\;  \mapsto  \;\;\;\left[\opr{H}_{\gamma},\opr{T}_{\gamma}\right]\,\subset \,i\, \opr{I}_{\gamma},\label{3rd}
\end{eqnarray}
where $\opr{I}_{\gamma}$ is the identity in the closure of the canonical domain $\mcal{D}_{can}^{\gamma}$. In most cases of quantum canonical pairs, the commutator and the canonical domains coincide and the canonical domain is dense. Because these domains do not coincide and that the canonical domain being non-dense for the pair $(\opr{H}_{\gamma},\opr{T}_{\gamma})$, one may question whether $\opr{H}_{\gamma}$ and $\opr{T}_{\gamma}$ can be appropriately labeled as a quantum canonical pair. A detailed answer to this issue has already been given by one of us in Reference-\cite{galapon3}, with which we refer the reader to.

It is sufficient to point out here that the canonical commutation relation (CCR) $[\opr{Q},\opr{P}]{\varphi}=i\hbar{\varphi}$ possesses numerous non-unitary equivalent solutions in a separable Hilbert space, and one such solution is the pair $(\opr{H}_{\gamma},\opr{T}_{\gamma})$. The set of properties of a specific solution is consequent to a set of underlying fundamental properties of the system under consideration or to the basic definitions of the operators involved or to some fundamental axioms of the theory or to some postulated properties of the physical universe, so that there is no preferred solution to the CCR. 

\section{Symmetries of the TOA-Operators and Relations among their Eigenfunctions}\label{sym}

In this section we derive the symmetries of the confined time of arrival operators, and from these symmetries we derive the basic properties of their eigenfunctions and eigenvalues. And from these symmetries we will infer the relationships among the eigenfunctions and eigenvalues for different values of the boundary parameter $\gamma$. Most important is the identification of these symmetries as analogues of the classical symmetries of the classical time of arrival.

Central to our discussion are the behaviors of the time of arrival operators under parity, $\Pi$, and under time reversal, $\Theta$, operations. Both operators are bounded and act on all vectors of the Hilbert space with the following corresponding operations in coordinate representation $\Pi\varphi(q,t)=\varphi(-q,t)$ and $\Theta\varphi(q,t)=\varphi^{*}(q,-t)$, respectively. In momentum representation, the actions of the parity and the time reversal operator are $\Pi\varphi(k,t)=\varphi(-k,t)$ and $\Theta\varphi(k,t)=\varphi^{*}(-k,-t)$, respectively. In the following discussions, we will only consider the vectors at $t=0$ so that reference to the parametric time $t$ can be omitted. 

\subsection{Non-Periodic $\gamma\neq\frac{\pi}{2}$ Case}

The symmetries of the non-periodic TOA-operators follow directly from the invariance of their kernel under the following operations, 
\begin{eqnarray}
\left<q\right|\opr{T}_{\gamma}\left|q'\right>&=&-\left<-q\right|\opr{T}_{\gamma}\left|-q'\right>^{*}\label{s1}\\
\left<q\right|\opr{T}_{\gamma}\left|q'\right>&=&-\left<q\right|\opr{T}_{-\gamma}\left|q'\right>^{*}\label{s2}\\
\left<q\right|\opr{T}_{\gamma}\left|q'\right>&=&\left<-q\right|\opr{T}_{-\gamma}\left|-q'\right>\label{s3}
\end{eqnarray}
We will find below that the above properties of the kernel dictates the properties of the eigenfunctions and eigenvalues of the $\opr{T_{\gamma}}$'s among themselves.

\subsubsection{Symmetry-1}
Let us derive the symmetry arising from equation-(\ref{s1}). Let $\varphi$ be any vector in the domain of $\opr{T}_{\gamma}$, which is the entire Hilbert space, then
\begin{equation}
\left(\opr{T}_{\gamma}\varphi\right)\!(q)=\int_{-l}^{l}\left<q\right|\opr{T}_{\gamma}\left|q'\right>\varphi (q') dq'
\end{equation}
Acting both sides of this equation by the parity operator and changing variables in the integration by $q'\rightarrow -q'$, we arrive at
\begin{equation}
\left(\Pi\opr{T}_{\gamma}\varphi\right)\!(q)=\int_{-l}^{l}\left<-q\right|\opr{T}_{\gamma}\left|-q'\right>\left(\Pi\varphi\right)\! (q') dq'
\end{equation}
Acting both sides of this equation by the time reversal operator yields,
\begin{equation}
\left(\Theta\Pi\opr{T}_{\gamma}\varphi\right)\!(q)=\int_{-l}^{l}\left<-q\right|\opr{T}_{\gamma}\left|-q'\right>^*\left(\Theta\Pi\varphi\right)\! (q') dq'
\end{equation}
where the identity $(\Theta\Pi\varphi)\!(q)=\varphi^*(-q)$ has been used. Applying equation-(\ref{s1}) finally gives $\left(\opr{T_{\gamma}}\Theta\Pi\varphi\right)\!(q)=-\left(\Theta\Pi\opr{T_{\gamma}}\varphi\right)\!(q)$. Since this relation holds in the entire Hilbert space, we get the following combined parity and time reversal symmetry of the $\opr{T_{\gamma}}$,
\begin{equation}\label{sym1}
\Pi^{-1}\Theta^{-1}\opr{T_{\gamma}}\Theta\Pi=-\opr{T_{\gamma}}.
\end{equation}

From equation (\ref{sym1}) we can infer the relationship among the eigenvalues and eigenfunctions of $\opr{T_{\gamma}}$ for a fixed $\gamma$. We know that $\opr{T_{\gamma}}$ is self-adjoint and compact for a given $\gamma$, and thus the eigenvalues are real and countable, in particular, they are either positive or negative. Let $\varphi_{\sigma,\gamma}$ be an eigenfunction of $\opr{T_{\gamma}}$ with the corresponding eigenvalue $\tau_{\sigma,\gamma}\neq 0$, where $\sigma$ constitutes the collection of quantum numbers necessary in specifying the eigenfunctions of $\opr{T_{\gamma}}$. Using equation (\ref{sym1}), we have $\opr{T_{\gamma}}\varphi_{\sigma,\gamma}=- \Pi^{-1}\Theta^{-1}\opr{T_{\gamma}}\Theta\Pi\varphi_{\sigma,\gamma}$. Because $\varphi_{\sigma,\gamma}$ is an eigenfunction of $\opr{T_{\gamma}}$ with the eigenvalue $\tau_{\sigma,\gamma}$, we get the relationship $-\tau_{\sigma,\gamma}\varphi_{\sigma,\gamma}=\Pi^{-1}\Theta^{-1}\opr{T_{\gamma}}\Theta\Pi\varphi_{\sigma,\gamma}$. And this implies the eigenvalue relation
\begin{equation}\label{osym1}
\opr{T_{\gamma}}\Theta\Pi\varphi_{\sigma,\gamma}=-\tau_{\sigma,\gamma}\Theta\Pi\varphi_{\sigma,\gamma}.
\end{equation}
Thus $\Theta\Pi\varphi_{\sigma,\gamma}$ is an eigenfunction of $\opr{T_{\gamma}}$ with the  eigenvalue $-\tau_{\sigma,\gamma}$.  Since $\tau_{\sigma,\gamma}$ is not zero, the eigenvalues of $\varphi_{\sigma,\gamma}$ and $\Theta\Pi\varphi_{\sigma,\gamma}$ have equal magnitudes but with opposite signs. We arrived at this conclusion from the reality of the eigenvalue.

Thus we have identified one quantum number $s$ which takes on either $\pm 1$, indicating the sign of the eigenvalue. Later on we will find that $s$ is related with the direction of propagation of the eigenfunctions. We indicate $s$ by writing the eigenfunctions in the form $\varphi_{\sigma,\gamma}^{\pm}$, where the $(+)$-sign indicates that it corresponds to the positive eigenvalue, and the $(-)$-sign indicates that it corresponds to the negative eigenvalue. We can now write also their corresponding eigenvalues as $\tau_{\sigma,\gamma}^{\pm}$. In particular, we have the relationships 
\begin{equation}\label{symsym1}
\varphi_{\sigma,\gamma}^{-}=\Theta\Pi\varphi_{\sigma,\gamma}^{+}\; \;\; \tau_{\sigma,\gamma}^{-}=-\tau_{\sigma,\gamma}^{+}
\end{equation}
where $\sigma$ now constitutes the rest of quantum numbers less $s$. Thus for every $\gamma$ and $\sigma$, there corresponds two eigenfunctions $\varphi_{\sigma,\gamma}^{\pm}$ which are related according to equation (\ref{symsym1}). 

In position and momentum representations, equation (\ref{symsym1}) leads to the eigenfunction relationships $\varphi_{\sigma,\gamma}^{-}(q)=\varphi_{\sigma,\gamma}^{+*}(-q)$ and $\varphi_{\sigma,\gamma}^{-}(k)=\varphi_{\sigma,\gamma}^{*}(k)$, where $\varphi_{\sigma,\gamma}^{\pm}(k)=\int_{-l}^{l}\varphi_{k}^{(\gamma)*}(q) \varphi_{\sigma,\gamma}^{\pm}(q)\, dq$, in which $k$ takes the discrete values $k=0,\, \pm1,\, \pm2, \dots$. And these lead to the probability density relations
\begin{equation}\label{prob1}
\abs{\varphi_{\sigma,\gamma}^{+}(q)}^2=\abs{\varphi_{\sigma,\gamma}^{-}(-q)}^2,\;\;\;
\abs{\varphi_{\sigma,\gamma}^{+}(k)}^2=\abs{\varphi_{\sigma,\gamma}^{-}(k)}^2.
\end{equation}
Equations (\ref{prob1}) mean that the position distributions corresponding to $\varphi_{\sigma,\gamma}^{-}$ and $\varphi_{\sigma,\gamma}^{+}$ are mirror images of each other, and the momentum distributions corresponding to the same eigenfunctions overlap.

\subsubsection{Symmetry-2}
Also using symmetry (\ref{s2}) one can show that $\left(\Theta\opr{T_{\gamma}}\varphi\right)\!(q)=-\left(\opr{T_{-\gamma}}\Theta\varphi\right)\!(q)$ for all $\varphi$ in the Hilbert space. This leads to the symmetry relation
\begin{equation}\label{sym2}
\Theta^{-1}\opr{T_{-\gamma}}\Theta=-\opr{T_{\gamma}}.
\end{equation}
Now let $\varphi_{\sigma,\gamma}^{\pm}$ be the eigenfunctions of $\opr{T_{\gamma}}$ for a given $\sigma$. 
%From the relation (\ref{sym2}), we have $\Theta^{-1}\opr{T_{-\gamma}}\Theta\varphi_{\sigma,\gamma}^{\pm}=-\opr{T_{\gamma}}\varphi_{\sigma,\gamma}^{\pm}=-\tau_{[\gamma]}^{\pm}\varphi_{\sigma,\gamma}^{\pm}$, which implies the eigenvalue equation
%\begin{equation}  
%\opr{T_{-\gamma}}\Theta\varphi_{\sigma,\gamma}^{\pm}=-\tau_{\sigma,\gamma}^{\pm}\Theta
%\varphi_{\sigma,\gamma}^{\pm}.
%\end{equation}
%Thus $\Theta\varphi_{\sigma,\gamma}^{\pm}$ is an eigenfunction of $\opr{T_{-\gamma}}$, with eigenvalue $-\tau_{\sigma,\gamma}^{\pm}$. 
Then, with our established notation above, we have the following relationship
\begin{equation}\label{sym2sym2}
\varphi_{\sigma,-\gamma}^{\mp}=\Theta\varphi_{\sigma,\gamma}^{\pm},\;\;\; \tau_{\sigma,-\gamma}^{\pm}=-\tau_{\sigma,\gamma}^{\mp}.
\end{equation}
In position and momentum representations, equation (\ref{sym2sym2}) leads to the eigenfunction relations $\varphi_{\sigma,-\gamma]}^{\mp}(q)=\varphi_{\sigma,\gamma}^{\pm*}(q)$ and $\varphi_{\sigma,-\gamma}^{\mp}(k)=\varphi_{\sigma,\gamma}^{\pm*}(-k)$. 
And these lead to the probability density relations
\begin{equation}\label{prob2}
\abs{\varphi_{\sigma,-\gamma}^{+}(q)}^2=\abs{\varphi_{\sigma,\gamma}^{-}(q)}^2,\;\;\;
\abs{\varphi_{\sigma,-\gamma}^{+}(k)}^2=\abs{\varphi_{\sigma,\gamma}^{-}(-k)}^2.
\end{equation}
That is, the position distributions corresponding to $\varphi_{\sigma,-\gamma}^{+}$ and $\varphi_{\sigma,\gamma}^{-}$ overlap; and the momentum distributions corresponding to $\varphi_{\sigma,\gamma}^{+}$ and $\varphi_{\sigma,\gamma}^{-}$ are mirror images of each other.

\subsubsection{Symmetry-3}
Using symmetry (\ref{s3}) it can be shown that $\left(\opr{T_{\gamma}}\Pi\varphi\right)\!(q)=\left(\Pi\opr{T_{\gamma}}\varphi\right)\!(q)$ for all $\varphi$ in the Hilbert space. This implies the symmetry relation
\begin{equation}\label{sym3}
\Pi^{-1}\opr{T_{-\gamma}}\Pi=\opr{T_{\gamma}}.
\end{equation}
Given the eigenfunctions of $\opr{T_{\gamma}}$, $\varphi_{\sigma,\gamma}^{\pm}$, and the symmetry relation (\ref{sym3}), it can be shown the eigenfunctions of $\opr{T_{-\gamma}}$ are also given by
\begin{equation}\label{symsym3}
\varphi_{\sigma,-\gamma}^{\pm}=\Pi\varphi_{\sigma,\gamma}^{\pm},\;\;\; \tau_{\sigma,-\gamma}^{\pm}=\tau_{\sigma,\gamma}^{\pm}
\end{equation}
These give the eigenfunction relations $\varphi_{\sigma,-\gamma}^{\pm}(q)=\varphi_{\sigma,\gamma}^{\pm}(-q)$ and $\varphi_{\sigma,-\gamma}^{\pm}(k)=\varphi_{\sigma,\gamma}^{\pm}(-k)$. From this follows the probability relation
\begin{equation}\label{prob3}
\abs{\varphi_{\sigma,-\gamma}^{\pm}(q)}^2=\abs{\varphi_{\sigma,\gamma}^{\pm}(-q)}^2,\;\;\;
\abs{\varphi_{\sigma,-\gamma}^{\pm}(k)}^2=\abs{\varphi_{\sigma,\gamma}^{\pm}(-k)}^2
\end{equation}
These mean that the position and momentum distributions corresponding to $\varphi_{\sigma,-\gamma}$ and $\varphi_{\sigma,\gamma}$ with the same eigenvalues are mirror images of each others. 

Below the above symmetries and probability relationships will be identified with the different symmetries of the classical time of arrival.

\subsection{Non-Periodic $\gamma=\frac{\pi}{2}$ and Periodic $\gamma=0$ Cases}
For $\gamma=\frac{\pi}{2}$ and $\gamma=0$, we find similar behaviors. For both cases, the kernel $\left<q\right|\opr{T}_{\gamma}\left|q'\right>$ is invariant under the following operations,
\begin{eqnarray}
\left<q\right|\opr{T}_{\gamma}\left|q'\right>&=&-\left<q\right|\opr{T}_{\gamma}\left|q'\right>^{*}\label{ss1}\\
\left<q\right|\opr{T}_{\gamma}\left|q'\right>&=&\left<-q\right|\opr{T}_{\gamma}\left|-q'\right>\label{ss3}
\end{eqnarray}
These symmetries dictate the relationship among the eigenfunctions of the time of arrival operators for $\gamma=0,\,\frac{\pi}{2}$.

Following the same method employed above, equations (\ref{ss1}) and (\ref{ss3}) lead to the following symmetries of the time of arrival operators for $\gamma=0,\frac{\pi}{2}$,
\begin{eqnarray}
\Theta^{-1}\opr{T_{\gamma}}\Theta&=&-\opr{T_{\gamma}}\label{sss1}\\
\Pi^{-1}\opr{T_{\gamma}}\Pi&=&\opr{T_{\gamma}}\label{sss3}
\end{eqnarray}
Likewise using the same arguments used above, equation (\ref{sss1}) leads to a pair of eigenfunctions with equal magnitudes of eigenvalues but with opposite signs, i.e. $\varphi_{\sigma,\gamma}^{\pm}$ and $\tau_{\sigma,\gamma}^{\pm}$. In particular equation (\ref{sss1}) yields  the following relationships between the eigenfunctions corresponding to the positive and negative eigenfunctions,
\begin{equation}
\varphi_{\sigma,\gamma}^{-}=\Theta\varphi_{\sigma,\gamma}^{+},\;\;\; \tau_{\sigma,\gamma}^{-}=-\tau_{\sigma,\gamma}^{+}.  \label{e1}
\end{equation}
On the other, hand equation (\ref{sss3}) implies that the eigenfunctions $\varphi_{\sigma,\gamma}^{\pm}$ are likewise eigenfunctions of the parity operator with even parity.  Thus we have symmetric position and momentum distributions, i.e. 
\begin{equation}\label{prob4}
\abs{\varphi_{\sigma,\gamma}^{+}(q)}^2=\abs{\varphi_{\sigma,\gamma}^{-}(q)}^2,\;\;\;\; \abs{\varphi_{\sigma,\gamma}^{+}(k)}^2=\abs{\varphi_{\sigma,\gamma}^{-}(k)}^2
\end{equation} 
overlapping both in position and momentum representations and symmetric about the origin. These imply the position and momentum operators have zero expectation values for these two cases, which further imply that the eigenfunctions for these two cases correspond to the classically indeterminate case of vanishing initial position and momentum.

\subsection{The Classical and Quantum Symmetries}
Earlier we raised the question which of the many time of arrival operators to consider; in particular, which of the $\opr{T_{\gamma}}$ is the appropriate time of arrival operator. We assert that all should be taken into account. We note that the classical time of arrival operator satisfies the following symmetries: $t(q,p)=-t(-q,p)$, $t(q,p)=-t(q,-p)$, and $t(q,p)=t(-q,-p)$. Comparing these relationships with equations (\ref{prob1}), (\ref{prob2}), and (\ref{prob3}), we find that there is a perfect correspondences between these sets. In particular, we have the following correspondences
\begin{eqnarray}
t(q,p)=-t(-q,p)&\longleftrightarrow & \opr{T_{\gamma}}=-\Theta^{-1}\Pi^{-1}\opr{T_{\gamma}}\Pi\Theta,\\
t(q,p)=-t(q,-p) &\longleftrightarrow & \opr{T_{\gamma}}=-\Theta^{-1}\opr{T_{-\gamma}}\Theta,\\
t(q,p)=t(-q,-p) &\longleftrightarrow & \opr{T_{\gamma}}=\Pi^{-1}\opr{T_{-\gamma}}\Pi.
\end{eqnarray}
It is then clear that all values for $\gamma\neq 0$ should be accounted for in order to accommodate the entire symmetry of the classical time of arrival. Thus not a single value of $\gamma$ is a sufficient quantization of the confined classical time of arrival. Note that from the symmetry of the eigenfunctions for $\gamma=0,\pi/2$, the CTOA-operators $\opr{T}_{0}$ and $\opr{T}_{\frac{\pi}{2}}$ correspond to the classically indeterminate case for $q=0$ and $p=0$.

\section{The CTOA-Eigenvalue Problem}\label{eigprob}
In the previous section, we arrive at the symmetries of the confined time of arrival operators, and these symmetries connect the different eigenfunctions and their corresponding eigenvalues. But that is how far the symmetries can give us. In this section, we solve the eigenvalue problem for the CTOA-operators, which is Fredholm integral operator problem of the second type. Our method is to convert the integral equation into its (integro) differential form. We are going to exploit the symmetry properties that we have derived above. It will be sufficient for us to solve explicitly for the positive-eigenvalue-eigenfunctions, and from them derive their corresponding negative-eigenvalue-eigenfunctions via their symmetry relationship established in the previous section.

\subsection{Non-periodic Confined Time of Arrival Operators, $\gamma\neq 0$}

Our problem now is to solve for the eigenvalue problem $\opr{T}_{\gamma}\varphi_{\tau}=\tau\varphi_{\tau}$, for the eigenfunction $\varphi_{\tau}$ and the corresponding eigenvalue $\tau$,for positive $\tau$. The eigenvalue equation is explicitly given by
\begin{eqnarray}
\tau\varphi_{\tau}(q)&=&\int_{-l}^l \left<q\right|\opr{T}_{\gamma}\left|q'\right>\varphi_{\tau}(q')\, dq'\nonumber\\
&=&-\frac{\mu e^{i\gamma}}{4\hbar \sin\gamma}
\int_{-l}^{q}(q+q')\varphi_{\tau}(q')\, dq'\nonumber\\
& &-\frac{\mu e^{-i\gamma}}{4\hbar \sin\gamma}\int_{q}^{l}(q+q')\varphi_{\tau}(q')\, dq'.
\label{expli1}
\end{eqnarray}
Differentiating equation-(\ref{expli1}) twice using Leibniz rule of differentiating an integral, yields the following differential equation for the eigenfunction,
\begin{equation}
\frac{d^{2}\varphi_{\tau}(q)}{dq^{2}}+\frac{\mu iq}{\tau \hbar}\frac{d \varphi_{\tau}(q)}{dq}+\frac{3\mu i}{2\tau \hbar}\varphi_{\tau}(q)=0.
\label{diffnonperiodic}
\end{equation}
The eigenfunctions are distinguished among the solutions of this differential equation by extracting and imposing the boundary condition from equation-(\ref{expli1}) itself. Evaluating equation-(\ref{expli1}) at the boundaries yields the following integro-boundary conditions on the eigenfunctions
\begin{eqnarray}
\varphi_{\tau}(l)\!\!&=&\!\!-\frac{\mu \,e^{i\gamma}}{4\tau\hbar \sin\gamma}
\int_{-l}^{l}(l+q') \varphi_{\tau}(q')\, dq',\label{bound1}\\
\varphi_{\tau}(-l)\!\!&=&\!\!-\frac{\mu \,e^{-i\gamma}}{4\tau\hbar \sin\gamma}
\int_{-l}^{l}(-l+q') \varphi_{\tau}(q')\, dq'.\label{bound2}
\end{eqnarray}
These are non-standard boundary conditions. Nevertheless they are sufficient to determine the eigenfunctions and eigenvalues.

\subsubsection{The general case}
Now we solve differential equation (\ref{diffnonperiodic}) for a given $\gamma$ subject to the conditions (\ref{bound1}) and (\ref{bound2}). Equation-(\ref{diffnonperiodic}) has a definite parity. If $\varphi(q)$ is a solution, then $\varphi(-q)$ is also solution. This can be seen by making the substitution $q\rightarrow -q$ in the differential equation. It is then sufficient for us to find odd and even solutions, and from them built the general solution by linear superposition. By power series method, we find the following odd and even solutions
\begin{eqnarray}
\varphi_{e}(q)=e^{-\frac{\mu iq^2}{4i\tau\hbar}}\left(\frac{\mu q^2}{\tau\hbar}\right)^
{\frac{3}{4}}\left[J_{-\frac{3}{4}}\!\!\left(\frac{\mu q^2}
{4\tau\hbar}\right)-iJ_{\frac{1}{4}}\!\!\left(\frac{\mu q^2}{4\tau\hbar}\right)\right]\label{evensol}
\end{eqnarray}
\begin{eqnarray}
\varphi_o(q)=q e^{-\frac{\mu iq^2}{4\tau\hbar}}
\left(\frac{\mu q^2}{\tau\hbar}\right)^{\frac{1}{4}}\!\!
\left[J_{-\frac{1}{4}}\!\!\left(\frac{\mu q^2}{4\tau\hbar}\right)
-iJ_{\frac{3}{4}}\!\!\left(\frac{\mu q^2}{4\tau\hbar}\right)\right]\label{oddsol}
\end{eqnarray}

The eigenfunctions are then of the form,
\begin{equation}
\varphi_{\tau}=A_0 \varphi_e(q)+A_1 \varphi_o(q)\label{gen}
\end{equation}
where $A_0$ and $A_1$ are constants yet to be determined from the boundary conditions.
Substituting equation-(\ref{gen}) back into both sides of equations (\ref{bound1}) and (\ref{bound2}), and  performing the indicated integrations, we obtain, after some simplification, the following system of equations for the unknown coefficients $A_0$ and $A_1$,
\begin{widetext}
\begin{eqnarray}
& &A_{0}\sqrt{\frac{\mu l^2}{\tau\hbar}}\left(J_{-\frac{3}{4}}\left(\frac{\mu l^2}{4\tau\hbar}\right)+\frac{1}{\tan  \gamma}J_{\frac{1}{4}}\left(\frac{\mu l^2}{4\tau\hbar}\right)\right)
+A_{1}l \left(J_{-\frac{1}{4}}\left(\frac{\mu l^2}{4\tau\hbar}\right)+\frac{1}{\tan \gamma}J_{\frac{3}{4}}\left(\frac{\mu l^2}{4\tau\hbar}\right)\right)=0\nonumber\\
& &A_{0}\sqrt{\frac{\mu l^2}{\tau\hbar}}\left(J_{-\frac{3}{4}}\left(\frac{\mu l^2}{4\tau\hbar}\right)-\frac{1}{\tan  \gamma}J_{\frac{1}{4}}\left(\frac{\mu l^2}{4\tau\hbar}\right)\right)
-A_{1}l \left(J_{-\frac{1}{4}}\left(\frac{\mu l^2}{4\tau\hbar}\right)-\frac{1}{\tan \gamma}J_{\frac{3}{4}}\left(\frac{\mu l^2}{4\tau\hbar}\right)\right)=0\nonumber
\label{eigencondition0}
\end{eqnarray}
This system of equations can be written in matrix form:
\begin{equation}
\left(
\begin{array}{rr}
\frac{\pi }{2\Gamma (\frac{3}{4})}\sqrt{\frac{\mu l^2}{\tau\hbar}}\left(J_{\frac{-3}{4}}\left(\frac{\mu l^2}{4\tau\hbar}\right)+\cot\gamma J_{\frac{1}{4}}\left(\frac{\mu l^2}{4\tau\hbar}\right)\right) & 
\Gamma(\frac{3}{4})\left(J_{\frac{-1}{4}}\left(\frac{\mu l^2}{4\tau\hbar}\right)+\cot\gamma J_{\frac{3}{4}}\left(\frac{\mu l^2}{4\tau\hbar}\right)\right)\\
\frac{\pi }{2\Gamma (\frac{3}{4})}\sqrt{\frac{\mu l^2}{\tau\hbar}}\left(J_{\frac{-3}{4}}\left(\frac{\mu l^2}{4\tau\hbar}\right)-\cot\gamma J_{\frac{1}{4}}\left(\frac{\mu l^2}{4\tau\hbar}\right)\right) &
\Gamma(\frac{3}{4})\left(\cot\gamma J_{\frac{3}{4}}\left(\frac{\mu l^2}{4\tau\hbar}\right)-J_{\frac{-1}{4}}\left(\frac{\mu l^2}{4\tau\hbar}\right)\right)\\
\end{array}
\right)
\left(
\begin{array}{r}
A_{0}\\
\\
A_{1}l\\
\end{array}
\right)
=0\\
\\
\label{thematrix}\nonumber
\end{equation}
\end{widetext}
In order for a non-trivial solution to exist, the determinant of the matrix of the coefficients of $A_0$ and $l A_1$ must vanish. The vanishing determinant leads to the condition
%\begin{equation}
%J_{-\frac{3}{4}}\!\!\left(\frac{\mu l^{2}}{4\tau\hbar}\right)J_{-\frac{1}{4}}\!\!\left(\frac{\mu l^{2}}{4\tau\hbar}\right)-\cot^2\gamma J_{\frac{3}{4}}\!\!\left(\frac{\mu l^{2}}{4\tau\hbar}\right)J_{\frac{1}{4}}\!\!\left(\frac{\mu l^{2}}{4\tau\hbar}\right)=0\nonumber
%\end{equation} 
%If we let $x=\mu l^{2}/4\tau\hbar$, the eigenvalue problem reduces to finding the roots of the transcendental equation
\begin{equation}
J_{-\frac{3}{4}}\!\!\left(x\right)J_{-\frac{1}{4}}\!\!\left(x\right)-\cot^2\gamma J_{\frac{3}{4}}\!\!\left(x\right)J_{\frac{1}{4}}\!\!\left(x\right)=0.\label{trans}
\end{equation} 
where $x=\mu l^{2}/4\tau\hbar$. By functional analytic arguments, the roots of this equation must exist and real and countably many. 

If we order the roots of equation-(\ref{trans}) such that $n=1$ corresponds to the first positive root $r_1$, $n=2$ to the second root $r_2$, and $n=m$ to the $m$-th root $r_m$, then we find that the remaining quantum number completely specifying the eigenfunctions of the confined non-periodic time of arrival operators consists of the positive integers $n$ ordering the roots of the equation-(\ref{trans}). For a given root $r_n$, we find that the constant $A_1$ has the form 
\begin{equation}
A_{1}=\frac{2\sqrt{r_{n}}}{l}\left(\frac{J_{-\frac{3}{4}}(r_{n})-\cot\gamma J_{\frac{1}{4}}(r_{n})}{J_{\frac{-1}{4}}(r_{n})-\cot\gamma J_{\frac{3}{4}}(r_{n})}\right) A_0
\end{equation}
Substituting $A_1$ back into equation-(\ref{gen}) gives the positive eigenvalue eigenfunction $\varphi_{n,\gamma}^+(q)$. 

And given $\varphi_{n,\gamma}^+(q)$ we can determine $\varphi_{n,\gamma}^-(q)$ from the symmetry relation (\ref{symsym1}), i.e. $\varphi_{n,\gamma}^-(q)=\Theta\Pi\varphi_{n,\gamma}^+(q)$. After performing some simplifications, the eigenfunctions are given by
\begin{widetext}
\begin{eqnarray}
\varphi_{n,\gamma}^{\pm}(q)&=&A_{n,\gamma} e^{\mp ir_n\frac{q^2}{l^2 }} \left(r_n\frac{q^2}{l^2}\right)^
{\frac{3}{4}}\left[J_{-\frac{3}{4}}\!\!\left( r_n\frac{q^2}{l^2}\right) \mp iJ_{\frac{1}{4}}\!\!\left( r_n\frac{q^2}{l^2}\right)\right] \left(J_{-\frac{1}{4}}\!(r_n)\!-\!\cot\gamma J_{\frac{3}{4}}\!(r_n)\right)\nonumber\\
& & \pm A_{n,\gamma} \frac{2 q \sqrt{r_n}}{l} e^{\mp ir_n\frac{q^2}{l^2 }} \left( r_n\frac{q^2}{l^2}\right)^{\frac{1}{4}} \left[J_{-\frac{1}{4}}\!\!\left( r_n\frac{q^2}{l^2}\right)
\mp iJ_{\frac{3}{4}}\!\!\left( r_n\frac{q^2}{l^2}\right)\right] \left({J_{-\frac{3}{4}}\!(r_n)-\cot\gamma J_{\frac{1}{4}}\!(r_n)}\right),\label{xxx}
\end{eqnarray}
\end{widetext}
where $A_{n,\gamma}$ is the normalization constant. The corresponding eigenvalues are 
\begin{equation}
\tau_{n,\gamma}^{\pm}=\pm \frac{l^2}{4\hbar}\,\frac{1}{r_n}.
\end{equation}
We shall call those that do not vanish elsewhere in the interval $[-l,l]$ as nonnodal-eigenfunctions; otherwise, as nodal eigenfunctions. The non-nodal (nodal) eigenfunctions correspond to those with even (odd) quantum number $n$. 

\subsubsection{The special anti-symmetric case, $\gamma\neq 0$}

The eigenfunctions for $\opr{T}_{\frac{\pi}{2}}$ can be derived directly from equation-({\ref{xxx}). For $\gamma=\pi/2$, equation-(\ref{trans}) reduces to $J_{-\frac{3}{4}}\!\!\left(x\right)J_{-\frac{1}{4}}\!\!\left(x\right)=0$. Then either $J_{-\frac{3}{4}}\!\!\left(x\right)=0$ or $J_{-\frac{1}{4}}\!\!\left(x\right)=0$. For the later the second term of the eigenfunction given by equation-(\ref{xxx}) vanishes; for the former, on the other hand, the first term vanishes. In this case, the eigenfunctions bifurcate into odd and even eigenfunctions.  The even and non-nodal eigenfunctions are 
%$J_{\nu,\rho}^{\mp}(x)=x^{\nu}(J_{-\nu}(x)\mp i J_{\rho}(x))$, and $A_0$ is the normalization constant.
\begin{equation}
\varphi_{s, \frac{\pi}{2},e}^{\pm}(q)\!=\!A_{s,\frac{\pi}{2}}^e e^{\mp ir_s\frac{q^2}{l^2 }}\!\! \left(r_s\frac{q^2}{l^2}\right)^
{\frac{3}{4}}\!\!\left[J_{-\frac{3}{4}}\!\!\left( r_s\frac{q^2}{l^2}\right)\!\!\mp \!\!iJ_{\frac{1}{4}}\!\!\left( r_s\frac{q^2}{l^2}\right)\right]
\label{eigenpi2even}
\end{equation}
with the eigenvalues given by $\tau_{s,\frac{\pi}{2},e}^{\pm}=\pm(\mu l^2/4\hbar)r_s^{-1}$, and the $r_s$'s are the positive roots of $J_{-\frac{3}{4}}(x)=0$, with $s=1, 2, \dots$. And the odd and nodal eigenfunctions are  \begin{equation}
\varphi_{u, \frac{\pi}{2},o}^{\pm}(q)\!\!=\!\!A_{u,\frac{\pi}{2}}^o q e^{\mp ir_u\frac{q^2}{l^2 }}\!\! \left( r_n\frac{q^2}{l^2}\right)^{\frac{1}{4}}\!\!\! \left[J_{-\frac{1}{4}}\!\!\!\left( r_u\frac{q^2}{l^2}\right)
\!\!\mp \!\!iJ_{\frac{3}{4}}\!\!\!\left(r_u\frac{q^2}{l^2}\right)\right]
\label{eigenpi2odd}
\end{equation}
with the eigenvalues given by $\tau^{\pm}_{u,\frac{\pi}{2},o}=\pm (\mu l^2/4\hbar)r_u^{-1}$, and the $r_u$'s are the positive roots of $J_{-\frac{1}{4}}(x)=0$, with $u=1,2,\dots$. $A_{s,\gamma}^e$ and $A_{u,\gamma}^o$ are normalization constants.

\subsection{Periodic CTOA-Operator, $\gamma=0$}
We turn to the eigenvalue problem $\opr{T}_0\varphi_{\tau}=\tau\varphi_{\tau}$, for positive $\tau$. Explicitly, the eigenvalue equation is given by
\begin{eqnarray}
\tau\varphi_{\tau}\!\!(q)&=&\frac{\mu}{4i\hbar}
\int_{-l}^{q}\varphi_{\tau}(q+q')dq'-\frac{\mu}{4i\hbar}
\int_{q}^{l}\varphi_{\tau}(q+q')dq'\nonumber\\
& &-\frac{1}{l}\int_{-l}^{l}\varphi_{\tau}(q^{2}-q'^{2})dq'\label{eigenzero1}
\end{eqnarray}
Differentiating this equation twice and after performing some simplifications yield the integro-differential equation
\begin{eqnarray}
\frac{d^{2}\varphi_{\tau}}{dq^{2}}+\frac{\mu iq}{\tau_{n} \hbar}\frac{d \varphi_{\tau}}{dq}+\frac{3\mu i}{2\tau \hbar}\varphi_{\tau}(q)
=\frac{\mu i}{2\tau \hbar l}\int_{-l}^{l}\!\!\varphi_{\tau}(q')dq'
\label{integrodiffperiodic}
\end{eqnarray}
And the eigenfunctions are subject to the integro-boundary conditions
\begin{eqnarray}
\varphi_{\tau}(l)=\frac{\mu}{4i \tau \hbar}\!
\left[\int_{-l}^{l}\!\!(l\!+\!q')\varphi_{\tau}(q')dq'\!-\!\!\frac{1}{l}\!\int_{-l}^{l}\!\!(l^{2}\!-\!q'^{2})\varphi_{\tau}(q')dq'\right]\nonumber
\end{eqnarray}
\begin{eqnarray}
\varphi_{\tau}(-l)\!=\!\frac{\mu}{4i \tau \hbar}\!\!
\left[\!\int_{-l}^{l}\!\!\!(l\!-\!q')\varphi_{\tau}(q')dq'\!-\!\frac{1}{l}\!\!\int_{-l}^{l}\!\!\!(l^{2}\!-\!q'^{2})\varphi_{\tau}(q')dq'\!\right]\nonumber
\end{eqnarray}
where we have arrived at these by evaluating equation-(\ref{integrodiffperiodic}) at the boundaries.

For us to have an idea of the possible solutions of the integro-differential equation satisfying the above boundary conditions, we subtract and add the boundary conditions to yield
\begin{equation}\label{intbc1}
\varphi_{\tau}(l)-\varphi_{\tau}(-l)=\frac{\mu}{i 2 \tau \hbar}\int_{-l}^l q' \varphi_{\tau}(q')\, dq'.
\end{equation}
\begin{equation}\label{intbc2}
\varphi_{\tau}(l)+\varphi_{\tau}(-l)=\frac{\mu}{i 2 l \tau \hbar}\int_{-l}^l q'^2 \varphi_{\tau}(q')\, dq'
\end{equation}
In this form of the boundary conditions, it is evident that solutions are odd and even functions of position. The even (odd) solution, for example, trivially satisfies the first (second), while it must satisfy the nontrivial second (first) condition. This is in fact predicted by our symmetry analysis, where we found that the eigenfunctions eigenfunctions of the parity operator as well.

\subsubsection{The Odd Eigenfunctions}
For odd eigenfunctions the right hand side of equation-(\ref{integrodiffperiodic}) vanishes and we are left with the same differential equation that we have for the non-periodic case. We know already that the odd solutions are of the form, which are given by equation-(\ref{oddsol}). For this case the integro-boundary condition (\ref{intbc1}) reduces to
\begin{eqnarray}
\varphi_{0, o}(l)=-\frac{\mu i}{4\tau \hbar}\int_{-l}^{l}\varphi_{0}(q')q'dq'.
\end{eqnarray}
Substituting the odd solution given by equation-(\ref{oddsol}) into the right hand side of this boundary condition gives $J_{-\frac{1}{4}}(x)=0$ with $x=\mu l^2/4\tau\hbar$.
%\begin{equation}
%J_{-\frac{1}{4}}\left(\frac{\mu l^2}{4\tau\hbar}\right)=0,
%\label{eigencondition1}
%\end{equation}
%which determines the eigenvalues.

Now let $r_n$ be the $n$th positive root of $J_{-\frac{1}{4}}(x)$. Then the positive eigenvalue corresponding to this root is $\tau_{n}^{+}=(\mu l^{2}/4\hbar)r_n^{-1}$. Substituting this eigenvalue back to equation-(\ref{oddsol}) gives the corresponding odd positive-eigenvalue-eigenfunction $\varphi^+_{n,0,o}(q)$. Given $\varphi^+_{n,0,o}(q)$ we likewise have the negative-eigenvalue-eigenfunction $\varphi^-_{n,0,o}(q)$ using symmetry relation (\ref{e1}), i.e. $\varphi^-_{n,0,o}(q)=\Theta\varphi^+_{n,0,o}(q)$. Explicitly, the odd eigenfunctions are given by
\begin{equation}
\varphi_{n, 0, o}^{\pm}(q)\!\!=\!\!A_{n,0}^o q e^{\mp ir_u\frac{q^2}{l^2 }}\!\! \left( r_n\frac{q^2}{l^2}\right)^{\frac{1}{4}}\!\!\! \left[J_{-\frac{1}{4}}\!\!\!\left( r_u\frac{q^2}{l^2}\right)
\!\!\mp \!\!iJ_{\frac{3}{4}}\!\!\!\left(r_u\frac{q^2}{l^2}\right)\right]
\label{eigenzerooddsimp}
\end{equation}
with the corresponding eigenvalues
\begin{equation}
\tau_{n}^{\pm}=\pm \frac{\mu l^{2}}{4r_{n}\hbar},
\label{eigeneven}
\end{equation}
for $n=1, 2,\dots$. $A_{n,0}^o$ is the normalization constant.

\subsubsection{The Even Eigenfunctions}
For even eigenfunctions, the right hand side of  equation-(\ref{integrodiffperiodic}) does not vanish. Since the right hand side involves only an integral of the unknown eigenfunction, the integral can be equated to a constant A, which is to be determined. This reduces the integro-differential equation into the form
\begin{eqnarray}
\frac{d^{2}\varphi_{0}(q)}{dq^{2}}+\frac{\mu iq}{\tau_{n} \hbar}\frac{d \varphi_{0}(q)}{dq}+\frac{3\mu i}{2\tau \hbar}\varphi_{0}(q)=\frac{\mu i}{2\tau \hbar l}A,
\label{integrodiffperiodiceven}
\end{eqnarray}
with the solutions subject to the integro-boundary condition
\begin{equation}\label{intbc22}
\varphi_{\tau}(l)=\frac{\mu}{i 4 l \tau \hbar}\int_{-l}^l q'^2 \varphi_{\tau}(q')\, dq'.
\end{equation}
Equation-(\ref{integrodiffperiodiceven}) is a linear, second - order non homogeneous differential equation. To solve it, it is sufficient to find the general solution to its associated homogeneous differential equation, which happens to be just equation-\ref{diffnonperiodic}.
%\begin{equation}
%\frac{d^{2}\varphi_{0}(q)}{dq^{2}}+\frac{\mu iq}{\tau \hbar}\frac{d \varphi_{0}(q)}{dq}+\frac{3\mu i}{2\tau \hbar}\varphi_{0}(q)=0
%\label{assoceven}
%\end{equation}
If $\varphi_0^a(q)$ is the general solution to the associated differential equation, the general solution to (\ref{integrodiffperiodiceven}) is given by
\begin{equation}
\varphi^{(0)}(q)=\varphi_{0}^{a}(q)+\frac{1}{3l}A\label{eigeneven1}.
\end{equation}
The even eigenfunctions then are found by obtaining the even solutions of the associated differential equation.

But we know already the solution to equation-(\ref{diffnonperiodic}). Substituting the even solution (\ref{evensol}) back in equation (\ref{eigeneven1}) and evaluating the solution at the boundaries yield the constant
\begin{eqnarray}
A=12 l \exp \left(-\frac{\mu i l^2}{4 \tau \hbar}\right)\left(\frac{\mu l^{2}}{\tau \hbar}\right)^{-\frac{1}{4}} J_{\frac{1}{4}}\left(\frac{\mu l^{2}}{4 \tau \hbar}\right).
\label{assocasum}
\end{eqnarray}
Substituting (\ref{evensol}) and (\ref{assocasum}) into (\ref{eigeneven1}), we obtain the following analytic form of the solution to the integro-differential equation,
\begin{eqnarray}
\varphi_{0, e}(q)\!\!&=&\!\!
e^{-\frac{\mu i q^2}{4 \tau \hbar}}\left(\frac{\mu q^{2}}{\tau \hbar}\right)^{\frac{3}{4}}\left[J_{-\frac{3}{4}}\!\!\left(\frac{\mu q^{2}}{4 \tau \hbar}\right)
-iJ_{\frac{1}{4}}\!\!\left(\frac{\mu q^{2}}{4 \tau \hbar}\right)\right]\nonumber\\
& &+4 e^{-\frac{\mu i l^2}{4 \tau \hbar}}\left(\frac{\mu l^{2}}{\tau \hbar}\right)^{-\frac{1}{4}}J_{\frac{1}{4}}\!\!\left(\frac{\mu l^{2}}{4 \tau \hbar}\right)
\label{eigeneven2}
\end{eqnarray}
The eigenvalues are found by imposing the integro-boundary condition on these solutions. Substituting equation-(\ref{eigeneven2}) back into (\ref{intbc22}) leads to the equality
%\begin{equation}
%J_{-\frac{3}{4}}\left(\frac{\mu l^{2}}{4\tau \hbar}\right)+\frac{2}{3}J_{\frac{5}{4}}
%\left(\frac{\mu l^2}{4\tau \hbar}\right)+\left(\frac{4\tau \hbar}{\mu l^{2}}\right)
%J_{\frac{1}{4}}\left(\frac{\mu l^{2}}{4\tau \hbar}\right)=0.
%\label{eigenconditioneven}
%\end{equation}
\begin{equation}
J_{-\frac{3}{4}}(x)+\frac{2}{3}J_{\frac{5}{4}}(x)+\frac{1}{x}
J_{\frac{1}{4}}(x)=0.
\label{eigenconditioneven}
\end{equation}
where $x=\mu l^{2}/4\tau \hbar$. %This determines the eigenvalues of the time of operator. 

The eigenvalue problem is then reduced to finding the roots of equation-\ref{eigenconditioneven}. Let $r_{s}$ be the $s$th positive root of this equation, the roots being ordered according to increasing magnitude. The $s$-th positive eigenvalue is then given by $r^+_{s}=\frac{\mu l^{2}}{4\tau_{s}\hbar}$. Substituting this back into equation-(\ref{eigeneven2}) gives the corresponding eigenfunction. From these positive eigenvalue eigenvalue eigenfunctions we derive the negative eigenvalue eigenfunctions using symmetry-(\ref{e1}). The eigenfunctions are now explicitly given by
\begin{eqnarray}
\varphi_{s,0,e}^{\pm}(q)\!\!&=&\!\!A_{s,0}^e  e^{\mp i\frac{q^2}{l^2} r_s}\!\! \left(\frac{q^2}{l^2} r_s\right)^{\frac{3}{4}}\!\!\left[\!J_{-\frac{3}{4}}\!\!\left(\frac{q^2}{l^2} r_s\right)\!
\mp \!iJ_{\frac{1}{4}}\!\!\left(\frac{q^2}{l^2} r_s\right)\!\right]\nonumber\\
& &\hspace{9mm}+  \frac{4 A_{s,0}^e e^{\mp ir_n} J_{\frac{1}{4}}(r_s)}{(4 r_s)^{\frac{1}{4}}}
\end{eqnarray}
where $A_{s,0}$ is the normalization constant. The corresponding eigenvalues are
\begin{equation}
\tau^{\pm}_{s}=\pm\frac{\mu l^{2}}{4 r_{s}\hbar},s=0,\pm 1, \pm 2\cdots
\end{equation}

\section{Dynamics}\label{dynamics}

The question now arises as to how we should interpret the eigenfunctions and eigenvalues of the confined time of arrival operators. Standard quantum mechanics postulates that the eigenvalues of an observable are the results of measurements when that observable is subject to measurement. 
In this section, we will demonstrate numerically that the spectral properties of these operators acquire interpretation independent from the measurement postulate. Here we will find that the spectral properties are instead tied with the dynamics of the system.

From the symmetries of the CTOA-operators, it is evident that the negative eigenvalue-eigenfunctions have exactly the same dynamics as those of the positive eigenvalue-eigenfunctions in the time reversed direction. It is then sufficient for us to consider in detail the dynamical behaviors of the positive eigenvalue eigenfunctions. We will classify the evolution according to whether the eigenfunction concerned is an eigenfunction of parity operator or not. The parity eigenfunctions are those of the $\gamma=0, \pi/2$ cases, and non-parity eigenfunctions otherwise.

\subsection{The quantum equation of motion}

Our analysis is based on the numerical integration of the evolution law $\varphi(t)=e^{-i\opr{H}_{\gamma}t/\hbar} \varphi(0)$ using spectral decomposition method, which, in position representation, is explicitly given by
\begin{eqnarray}
\varphi(q,t)&=&\sum_{k=-\infty}^{\infty} \fprod{\phi_k^{\gamma}}{\varphi(0)} e^{-i E_k t/\hbar} \phi_k^{\gamma}(q)\nonumber\\
%&=&\sum_{k=-\infty}^{\infty} \fprod{\phi_k^{\gamma}}{\varphi(0)} e^{-i E_k t/\hbar} \frac{1}{\sqrt{2l}} e^{i(\gamma+k\pi)\frac{q}{l}}.
%\end{eqnarray}
%This sum can be rewritten in the form
%\begin{equation}\label{dyna}
%\varphi(q,t)
&=&e^{i\gamma \frac{q}{l}}\sum_{k=-\infty}^{\infty} b_k(t)\,\, \frac{e^{ik\pi\frac{q}{l}}}{\sqrt{2l}}.\label{dyna}
%\end{equation}
\end{eqnarray}
where $b_k(t)=e^{-i E_k t/\hbar}\int_{-l}^l \phi_k^{\gamma *}(q) \varphi(q,0) dq$. Evidently the time evolution is a reconstruction by Fourier series. 

Central to our analysis for the interpretation of the eigenfunctions and eigenvalues are the expectation value and the variance of the position operator as a function of time with respect to the eigenfunctions of the confined time of arrival operators. That is the quantities
\begin{equation}\label{ave}
\left<\opr{q}\right>_{n,\gamma}\!(t)=\int_{-l}^l q \abs{\varphi_{n,\gamma}\!(q,t)}^2\, dq
\end{equation}
\begin{equation}\label{var}
\sigma_{n,\gamma}^2\!(t)\!=\!\!\int_{-l}^l\!\! q^2 \abs{\varphi_{n,\gamma}\!(q,t)}^2\, dq-\left(\int_{-l}^l q \abs{\varphi_{n,\gamma}\!(q,t)}^2\, dq\right)^2
\end{equation}
where the $\varphi_{n,\gamma}(q,t)$'s are the evolved eigenfunctions of the confined time of arrival operators.

\subsection{Numerical evolution and the Gibbs phenomenon}
Since there is no closed form for the Fourier coefficients of the eigenfunctions, we resort to numerical evaluation of the coefficients and hence a numerical evaluation of the sum involved in the evaluation of equation-(\ref{dyna}). The numerical implementation of equation-(\ref{dyna}) is then a special case of the  general truncated Fourier series
\begin{equation}
f_N(x)=\sum_{k=-N}^N f_k \, e^{i k\pi x},
\end{equation}
where the $f_k$'s are the Fourier coefficients given by, in the rescaled interval $[-1,1]$, $f_k=\int_{-1}^1  e^{-i k\pi x} f(x) dx$. 

It is well known that when $f(x)$ is analytic and periodic the Fourier series converges exponentially fast, i.e. $\mbox{max}\abs{f(x)-f_N(x)}\leq e^{-\alpha N}$ for $-1<x<1$ for some $\alpha>0$. In such cases the truncated Fourier sum is an accurate approximation of $f(x)$ for sufficiently large $N$. However, when $f(x)$ is non-periodic and/or discontinuous functions the Fourier sum converges slowly in the interval $(-1,1)$, i.e. $\abs{f(x_0)-f_N(x_0)}\approx O(N^{-1})$ for $-1<x_0<1$. And there is an overshoot at the boundary that does not diminish with increasing $N$, i.e. $\mbox{max}\abs{f(x)-f_N(x)}$ for $-1<x<1$ does not tend to zero as $N$ increases indefinitely. This is the well-known  Gibbs phenomenon, which undermines obtaining accurate point values from the knowledge of Fourier coefficients for  non-periodic functions \cite{lieb}.

The presence of Gibbs phenomenon undermines our intention to understand the dynamics of the eigenfunctions. This can be seen from equations-(\ref{ave}) and-(\ref{var}) where we need accurate point values of the evolved eigenfunctions in order to get an accurate value of the required integrals. In order to have an accurate picture of the evolution of the expectation value and variance of the position operator, one must have first an accurate numerical approximation to the evolved eigenfunctions. Thus in the following we limit our numerical evolution to particular set of eigenfunctions and values of $\gamma$ in order to avoid Gibb's phenomenon.

\subsection{Results}

\subsubsection{Parity eigenfunctions}

The eigenfunctions for the symmetric and anti-symmetric confined time of arrival operators are parity eigenfunctions, being even and odd functions of position. For the $\gamma=0$ case we can only evolve directly the even eigenfunctions via Fourier series without the effect of Gibbs phenomenon. For the $\gamma=\pi/2$ case, we can evolve odd eigenfunctions without the effect of Gibbs phenomenon. That that is the case for the anti-periodic case can be seen from the following argument. The Fourier coefficients are given by 
\begin{equation}
\int_{-l}^l \phi_k^{\frac{\pi}{2} *}(q) \varphi_{n,\frac{\pi}{2}}(q) dq=\!\!\frac{1}{\sqrt{2l}}\int_{-l}^l\!\!\! e^{-i k \pi \frac{q}{l}} \left(e^{-i\pi \frac{q}{l}}\varphi_{n,\frac{\pi}{2}}(q)\right) dq\nonumber.
\end{equation}
Evidently the sum in equation-(\ref{dyna}) is the Fourier sum for the analytic function $\left(e^{-i\pi \frac{q}{l}}\varphi_{n,\frac{\pi}{2}}(q)\right)$. For even $\varphi_{n,\frac{\pi}{2}}(q)$ the function $\left(e^{-i\pi \frac{q}{l}}\varphi_{n,\frac{\pi}{2}}(q)\right)$ is odd, thus nonperiodic; but for odd eigenfunction, it is even, so that it is periodic.

%%%%%%%%%%%%%%%%%%%%Figure 1%%%%%%%%%%%%%%%%%%%%%%%%%%%%%%%
\begin{figure}[!tbp]
{\includegraphics[height=1.75in,width=3.5in,angle=0]{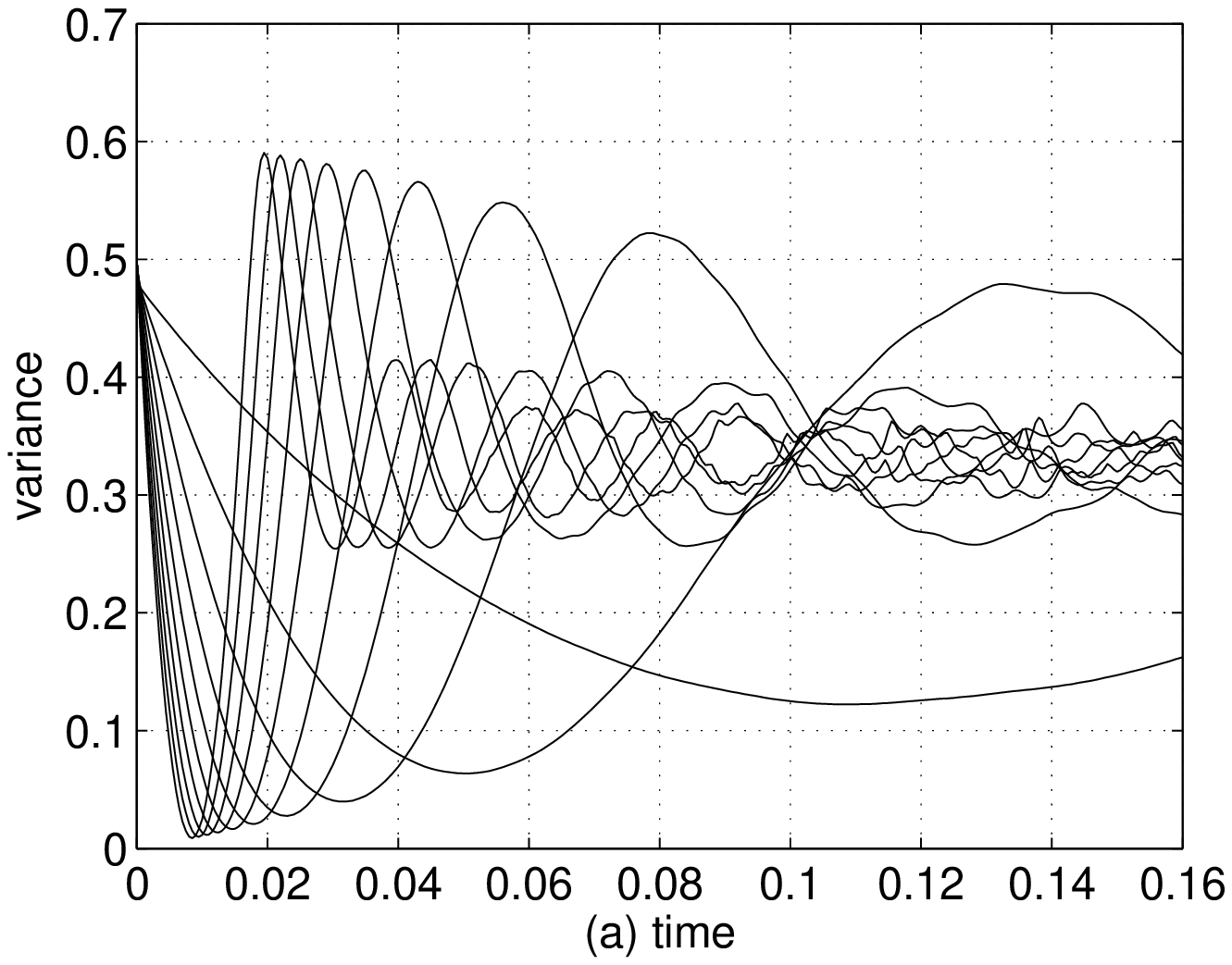}}
{\includegraphics[height=1.75in,width=3.5in,angle=0]{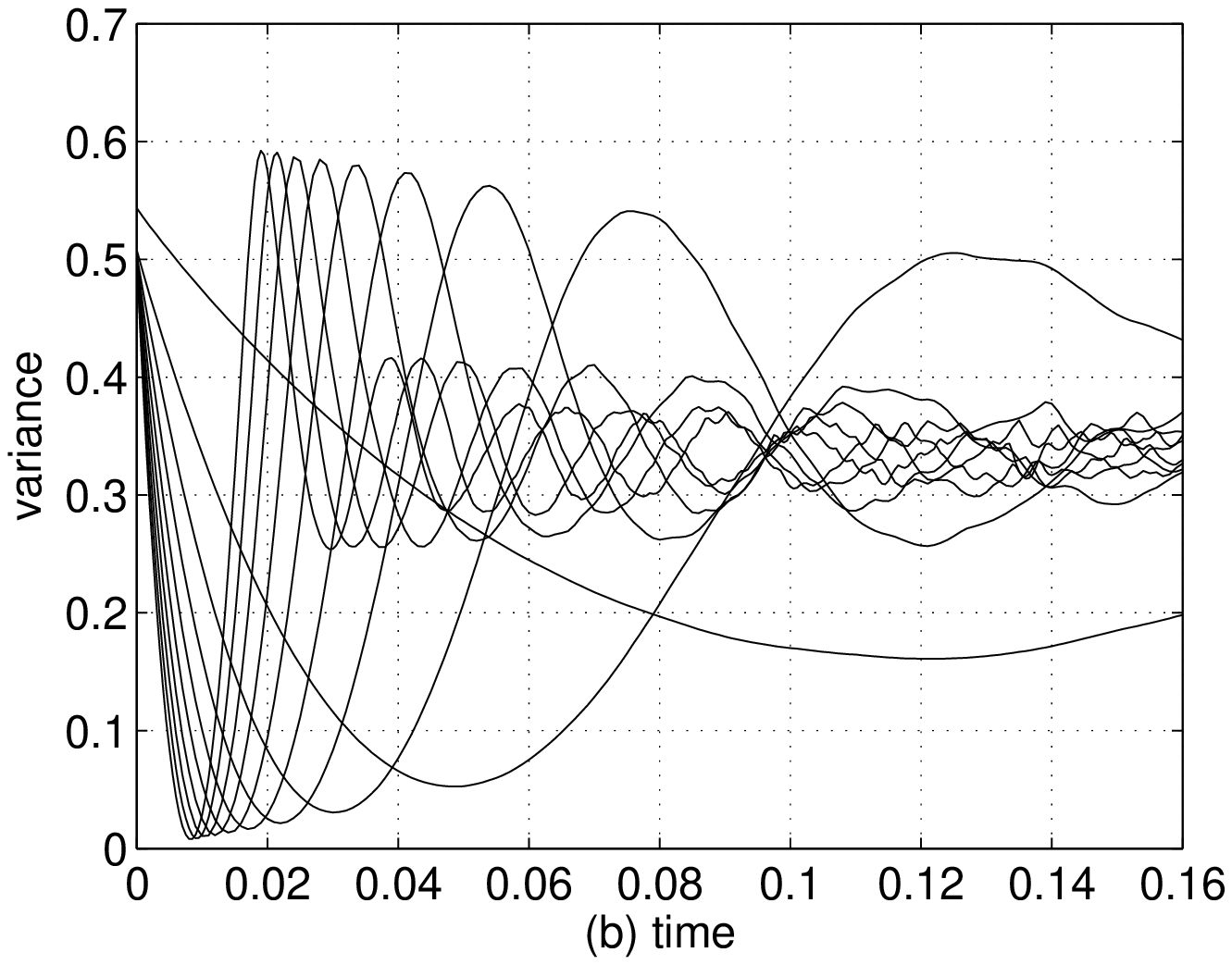}}
{\includegraphics[height=1.75in,width=3.5in,angle=0]{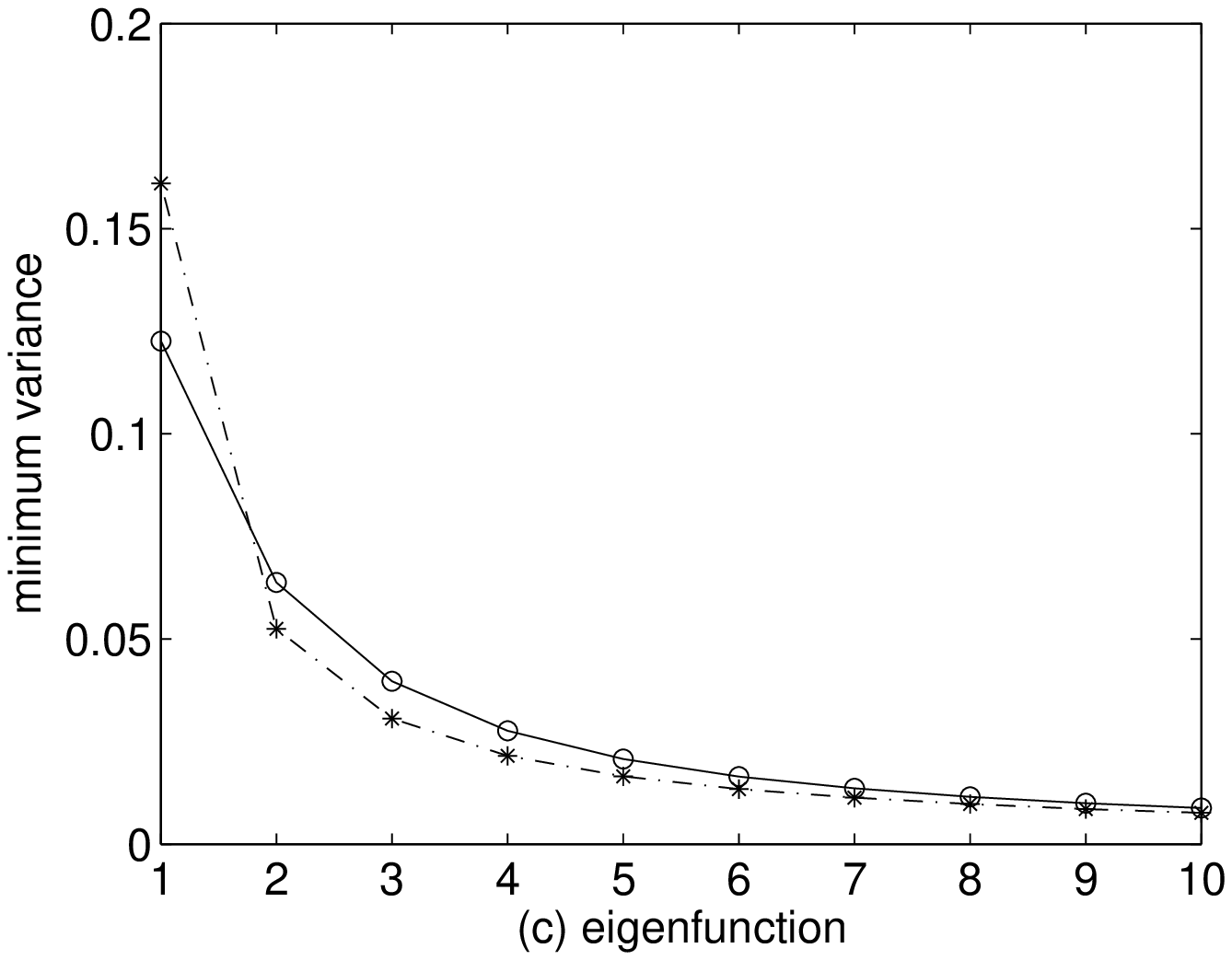}}
\caption{The variance as a function of time for the first ten symmetric CTOA-even-eigenfunctions for $\gamma=0$, ({\it a});
and antisymmetric CTOA-odd-eigenfunctions for $\gamma=\pi/2$, ({\it b}). The corresponding minimum variance for these eigenfunctions are shown in ({\it c}). $l=\mu=\hbar=1.$ All units, with all in succeding figures, are in atomic units.}\label{fig:varparity} 
\end{figure}
%%%%%%%%%%%%%%%%%%%%%%%%%%%%%%%%%%%%%%%%%%%%%%%%%%%%%%%%%%%%%%%%

Here we demonstrate numerically that these eigenfunctions unitarily arrive, i.e. the eigenfunctins evolve according to Schrodinger's equation, at the origin at their respective eigenvalues. These unitary arrival is measured by the variance of the position operator with respect to these eigenfunctions. Figure-\ref{fig:varparity} shows the variance as a function of time for the first ten largest eigenvalue-eigenfunctions for the even and odd periodic and anti-periodic CTOA-eigenfunctions, respectively. Starting from $t=0$ the variance decreases, reaches a minimum then increases again, followed by a decaying oscillation. The figure clearly demonstrates that the variance decreases with $n$, so that the eigenfunctions are arbitrarily localized at the origin for arbitrarily large $n$. Tables- I and-II shows the minimum variance for a given eigenfunction and the interpolated variance at the eigenfunction's corresponding eigenvalue. For the parameters indicated in the caption, the minimum variance and the interpolated variance at the eigenvalue agree at least to three significant figures. Our results shows that the eigenfunctions evolved such that the variance is minimum at their eigenvalues.

Figure-\ref{fig:evovar} shows the general features of the evolving probability density for even and odd parity eigenfunctions. The even parity eigenfunctions evolve such that their corresponding probability densities obtain their minimum widths, and their peaks being maximum at the origin at their respective eigenvalues. The probability densities for the odd eigenfunctions, on the other hand, evolve with two peaks approaching the origin, the value of the probability density being zero at the origin for all times. The time of closest approach to the origin of the two peaks occur at the eigenvalue. We shall refer to the former as non-nodal and the latter nodal. We will find below that the dynamical behaviors of these eigenfunctions are also shared by the eigenfunctions for the non-periodic CTO-operator eigenfunctions.

%\begin{widetext}
\begin{center}
\begin{table}[t]
\begin{center}
\begin{tabular}{c}\hline
Periodic Even CTOA-eigenfunctions\\
\begin{tabular}{|c|c|c|c|}\hline
$n$-th        & Eigenvalue         & Computed       & Interpolated   \\
even          & ($\times 10^{-3}$) & minimum          & variance at the            \\
eigen-        &                    & variance in the  & eigenvalue       \\
function      &                    & interval         &          \\ 
              &                    & ($\times 10^{-3}$)& ($\times 10^{-3}$)         \\ 
\hline\hline
1             & 111.43823   & 122.4                    & 122.6                       \\
\hline
2             &  50.17966   & 63.72                    & 63.76                       \\
\hline
3             & 31.58532    &  39.72                   & 39.72                        \\
\hline
4             & 22.84895    & 27.63                    & 27.63                        \\
\hline
5             & 17.84116    & 20.76                    & 20.76                       \\
\hline
6             & 14.61367    & 16.46                    & 16.46                       \\
\hline
7             & 12.36664    & 13.57                    & 13.57                       \\
\hline
8             & 10.71462    & 11.50                    & 11.50                       \\
\hline
9             & 9.44995     & 9.970                    & 9.971                       \\
\hline
10            & 8.45118     & 8.787                    & 8.787                       \\
\hline
\end{tabular}
\end{tabular}
\end{center}
\caption{Comparison of the calculated minimum variance with that of the linearly
interpolated variance at the eigenvalue for the corresponding eigenfunction for
the first ten even functions of the periodic CTOA-operator. The variances are
calculated with 401 Fourier coefficients at 0.0001 time steps.}
\end{table}
\end{center}
%\end{widetext}

\begin{center}
\begin{table}[t]
\begin{center}
\begin{tabular}{c}\hline
Anti-Periodic Odd CTOA-eigenfunctions\\
\begin{tabular}{|c|c|c|c|}\hline
$n$-th        & Eigenvalue         & Computed       & Interpolated               \\
odd           & ($\times 10^{-3}$) &   minimum        & variance at the            \\
eigen-        &                    & variance in the  & eigenvalue                  \\
function      &                    & interval         &                             \\ 
              &                    & ($\times 10^{-3}$)& ($\times 10^{-3}$)         \\ 
\hline\hline
1             & 124.60751   &   161.0                  &    161.3                    \\
\hline
2             & 48.79893   &    52.46                 &      52.47                  \\
\hline
3             & 30.27385    &    30.61                 &     30.61                   \\
\hline
4             & 21.93662    &    21.50                 &     21.50                   \\
\hline
5             &  17.19832   &    16.53                &       16.53                 \\
\hline
6             & 14.14287    &     13.42               &       13.43                 \\
\hline
7             & 12.00910    &     11.29                &     11.29                  \\
\hline
8             & 10.43469    &     9.741                &      9.746                 \\
\hline
9             & 9.22520     &      8.558              &        8.558               \\
\hline
10            &  8.26694    &      7.631              &        7.633               \\
\hline
\end{tabular}
\end{tabular}
\end{center}
\caption{Comparison of the calculated minimum variance with that of the linearly
interpolated variance at the eigenvalue for the corresponding eigenfunction for
the first ten odd functions of the anti-periodic CTOA-operator. The variances are
calculated with 401 Fourier coefficients at 0.0001 time steps, except for the 9-th and 10-th
eigenfunctions where the time step is 0.00005.}
\end{table}
\end{center}
%\end{widetext}
%%%%%%%%%%%%%%%%%%%%Figure 1%%%%%%%%%%%%%%%%%%%%%%%%%%%%%%%
\begin{figure}[!tbp]
{\includegraphics[height=1.5in,width=3.5in,angle=0]{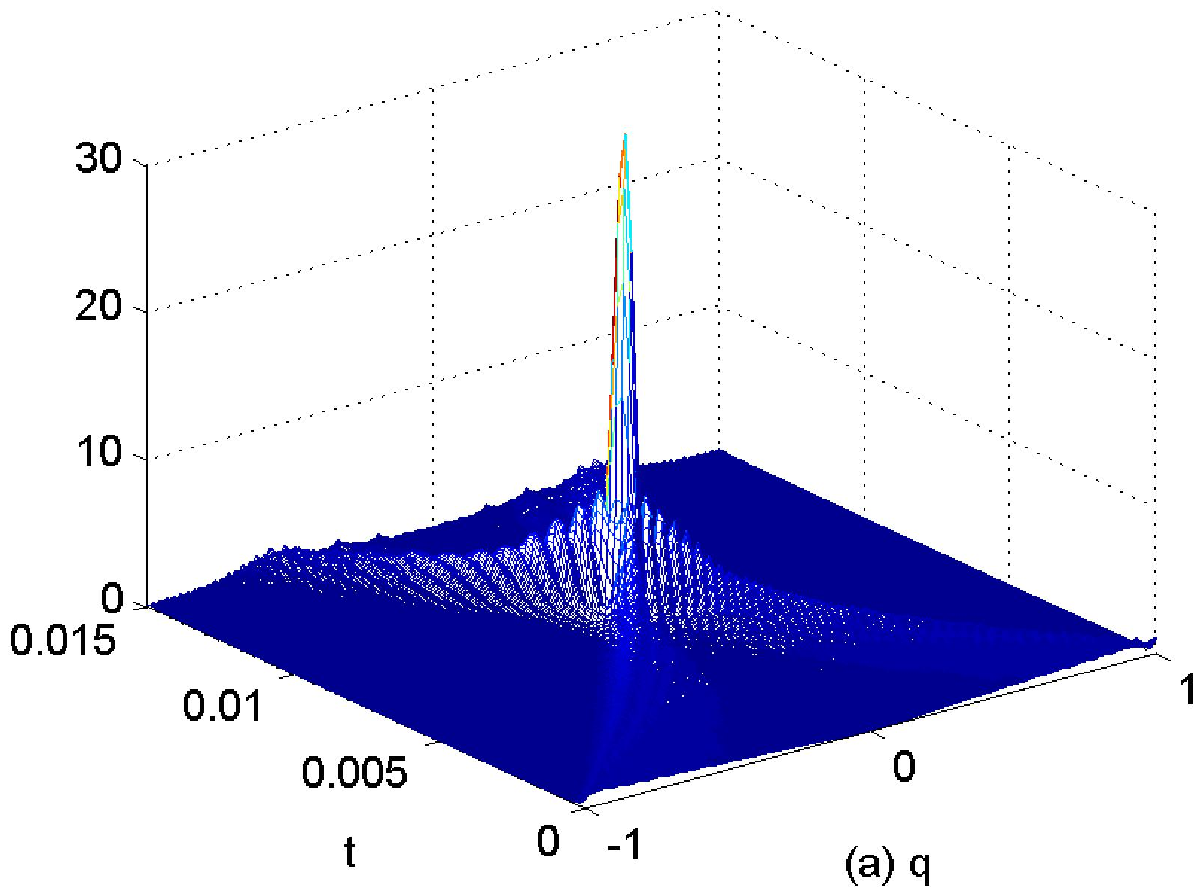}}
{\includegraphics[height=1.5in,width=3.5in,angle=0]{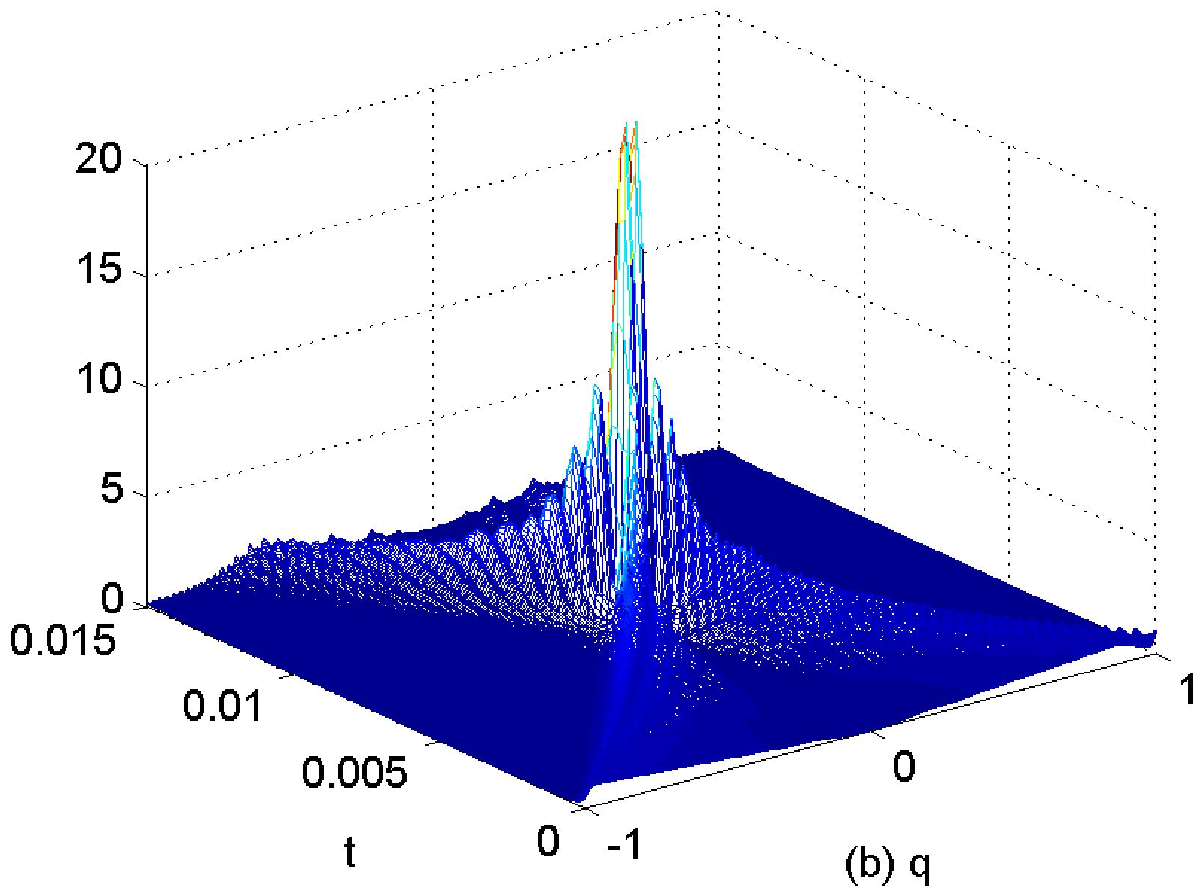}}
\caption{(Color online) The $n=10$ (a) evolved probability density for $\gamma=0$, and $n=10$ (b) evolved probability density for $\gamma=\pi/2$, with $\hbar=l=m=1$. Both symmetrically ``collapse'' at the origin at their respective eigenvalues.}\label{fig:evovar}
\end{figure}
%%%%%%%%%%%%%%%%%%%%%%%%%%%%%%%%%%%%%%%%%%%%%%%%%%%%%%%%%%%%%%%%

%%%%%%%%%%%%%%%%%%%%Figure 1%%%%%%%%%%%%%%%%%%%%%%%%%%%%%%%
%\begin{figure}[h]
%{\includegraphics[height=1.45in,width=3.25in,angle=0]{minimum_variance_evensym_oddansym.eps}}
%\caption{The variance as a function of time for the first ten symmetric CTOA-even-egenfunctions ({\it a})
%and antisymmetric CTOA-odd-eigenfunctions ({\it b}).}
%\end{figure}
%%%%%%%%%%%%%%%%%%%%%%%%%%%%%%%%%%%%%%%%%%%%%%%%%%%%%%%%%%%%%%%%

\subsubsection{Non-parity eigenfunctions}

The eigenfunctions for $\gamma\neq 0$ are non-parity eigenfunctions. These eigenfunctions are nonperiodic, hence subject to Gibbs phenomenon. However, we can choose $\gamma$ such that Gibbs phenomenon can be ``reduced''. Notice that the the eigenfunctions are linear superpositions of odd and even functions of $q$. The idea is to choose $\gamma$ such that the even part dominates the odd part. If this can be done, the eigenfunctions are approximately periodic, and hope that Gibbs phenomenon is not that ``large''. Fortunately, this can be done. Recall that the eigenvalues are determined from the roots of the characteristic equation $J_{-\frac{3}{4}}(x) J_{-\frac{1}{4}}(x)-\cot^2\gamma J_{\frac{3}{4}}(x) J_{\frac{1}{4}}(x)=0$. For sufficiently small $\gamma$, the second term of the characteristic equation dominates and the first term and the eigenvalue condition reduces to $J_{\frac{3}{4}}(x) J_{\frac{1}{4}}(x)\approx 0$. The eigenvalues are now approximated by the roots of $J_{\frac{3}{4}}(x)$ and $J_{\frac{1}{4}}(x)$. For the eigenfunctions corresponding to the roots of $J_{\frac{1}{4}}(x)$, the even term dominates the odd term, and the eigenfunctions are approximately periodic at the boundaries. On the other hand, for the eigenfunctions corresponding to the roots of $J_{\frac{3}{4}}(x)$, the eigenfunctions are non-periodic, and we expect that Gibbs phenomenon has considerable effect on the numerical sum. Figure-\ref{fig:comp} gives a graphical comparison of the The functions $J_{\frac{1}{4}}(x)$, $J_{\frac{3}{4}}(x)$, and $J_{-\frac{3}{4}}(x) J_{-\frac{1}{4}}(x)-\cot^2\gamma J_{\frac{3}{4}}(x) J_{\frac{1}{4}}(x)$ for $\gamma=0.01$. It shows that for the given $\gamma$ the roots of $J_{\frac{1}{4}}(x)$ and $J_{\frac{3}{4}}(x)$ approximate the roots of $J_{-\frac{3}{4}}(x) J_{-\frac{1}{4}}(x)-\cot^2\gamma J_{\frac{3}{4}}(x) J_{\frac{1}{4}}(x)$. 

%%%%%%%%%%%%%%%%%%%%Figure 1%%%%%%%%%%%%%%%%%%%%%%%%%%%%%%%
\begin{figure}[h]
{\includegraphics[height=2.25in,width=3.5in,angle=0]{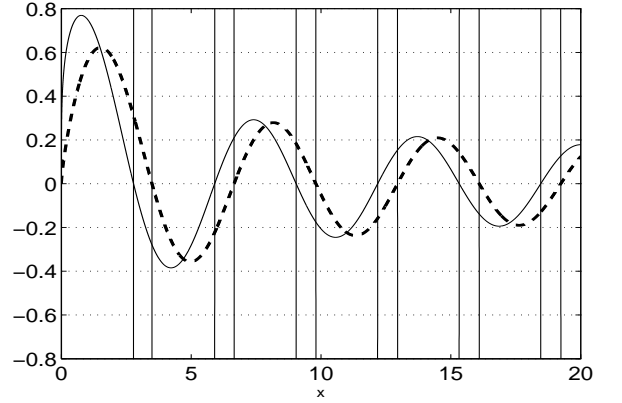}}
\caption{The functions $J_{\frac{1}{4}}(x)$ (solid line), $J_{\frac{3}{4}}(x)$ (dashed line), and $J_{-\frac{3}{4}}(x) J_{-\frac{1}{4}}(x)-\cot^2(0.01) J_{\frac{3}{4}}(x) J_{\frac{1}{4}}(x)$ (the almost vertical lines).} \label{fig:comp}
\end{figure}
%%%%%%%%%%%%%%%%%%%%%%%%%%%%%%%%%%%%%%%%%%%%%%%%%%%%%%%%%%%%%%%%

Figure-\ref{fig:asymvar}.$a$ shows the behavior of the expectation value of the position operator for each eigenfunction for $\gamma=0.01$. We find that the eigenfunctions for even $n$ approach the origin from the positive side axis; while the eigenfunctions for odd $n$, from the negative axis. Figure-\ref{fig:asymvar}$b$ shows the variance as a function of time. Evidently the variance has the same behavior as those of the parity eigenfunctions for $\gamma=\pi/2, 0$; and that the minimum variance decreases with $n$ so that the eigenfunctions become increasing localized at the origin for increasing $n$. The eigenfunctions are also either nodal or nonnodal with the same dynamical behaviors as those of the parity eigenfunctions. For the given $\gamma$, the nodals are those with $n$ odd; and non-nodals, with $n$ even. Figure-\ref{fig:asym} shows the general features of the probability density as a function of time and space for nodal and non-nodal eigenfunctions. Generally the zero of the nodal eigenfunctions do not occur at the origin as those of the nodal parity eigenfunctions; however, the zero approaches the origin in time and coincides with the origin at the eigenvalue.

Table-III summarizes the minimum variance, together with the linearly interpolated position expectation value and variance at the eigenvalue. It shows that at the eigenvalue, the expectation value is qualified zero for all cases. On other hand, the minimum variance and the interpolated variance at the eigenvalue match at least to three significant figures for $n$-even and at most two significant figures for $n$-odd. The former correspond to the case when the eigenfunction is approximately periodic at the boundary as discussed above; the latter, when the eigenfunction is non-periodic and Gibbs phenomenon is prevalent. The blank entries for the largest eigenvalue indicates the failure of our method to converge. This is due to the large oscillations in the exponentials for long times of evolution.

%%%%%%%%%%%%%%%%%%%%Figure 1%%%%%%%%%%%%%%%%%%%%%%%%%%%%%%%
\begin{figure}[!tbp]
{\includegraphics[height=1.47in,width=3.25in,angle=0]{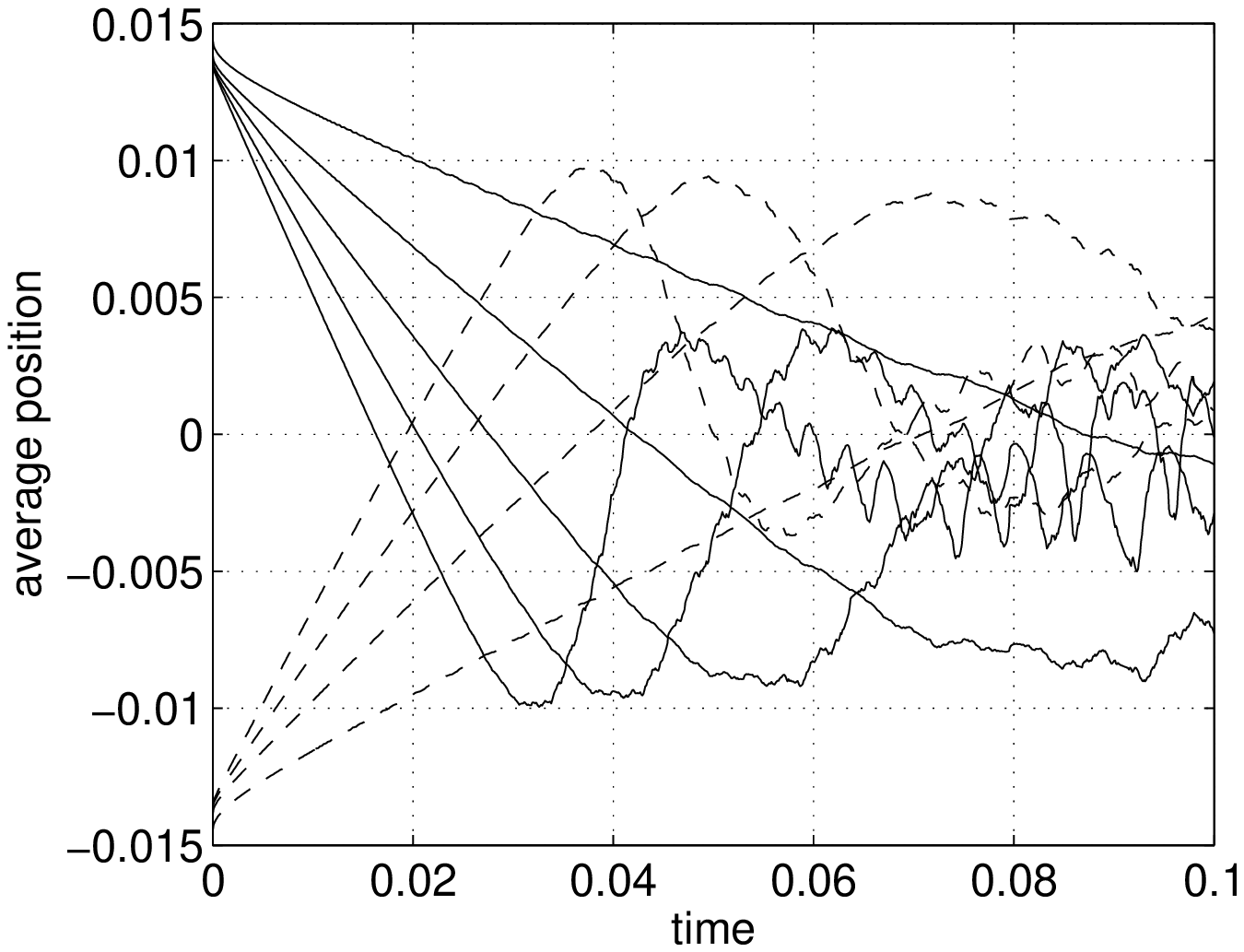}}
{\includegraphics[height=1.45in,width=3.25in,angle=0]{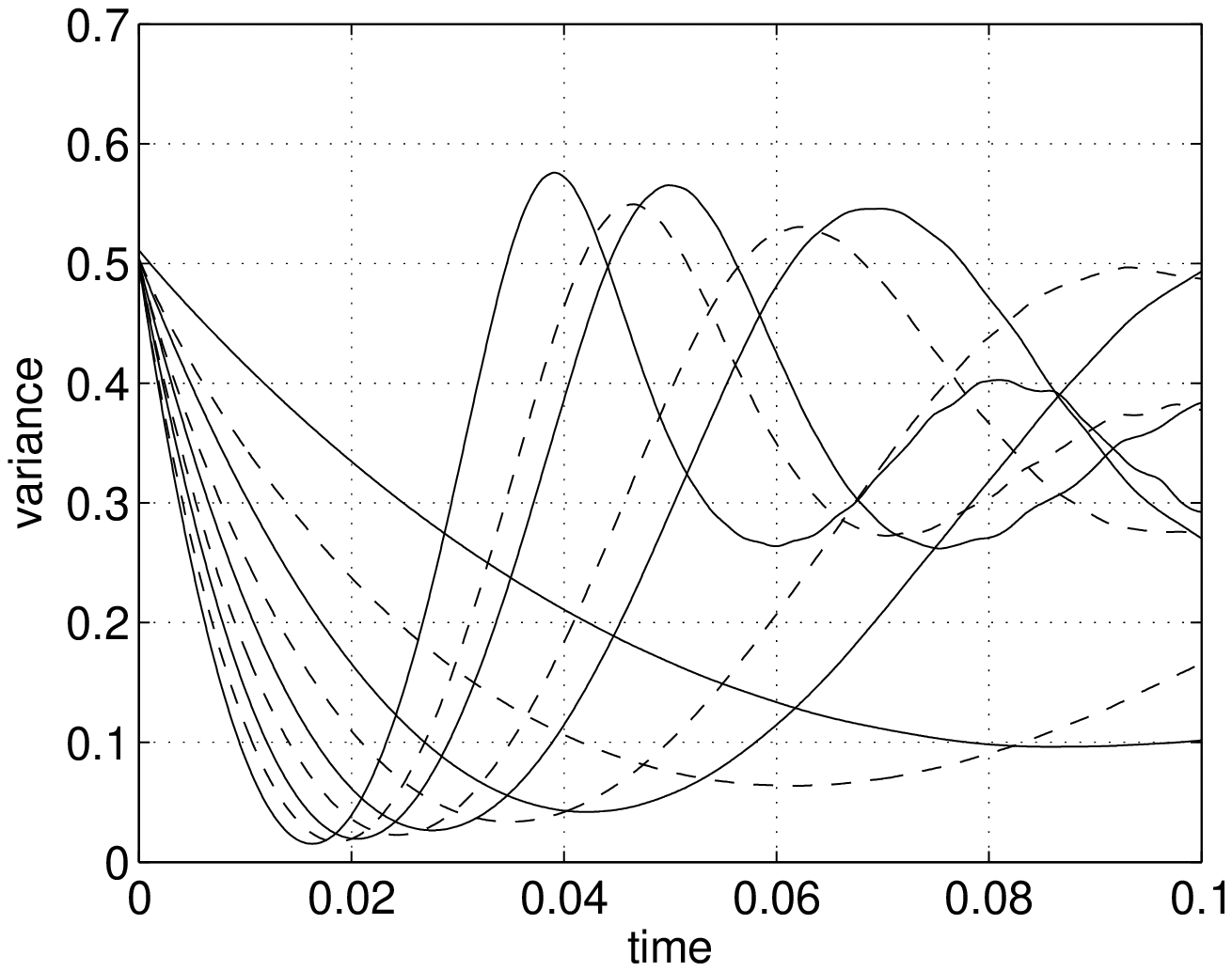}}
\caption{The top-figure shows the expectation value of the position operator for the first ten (10) eigenfunctions for $\gamma=0.01$. The solid line corresponds to $n$ equals even; dashed line, $n$ equals odd. The bottom-figure shows the variances, $\sigma^2(t)$, of the same eigenfunctions as a function of time. $l=\mu=\hbar=1.$}\label{fig:asymvar}
\end{figure}
%%%%%%%%%%%%%%%%%%%%%%%%%%%%%%%%%%%%%%%%%%%%%%%%%%%%%%%%%%%%%%%%

%%%%%%%%%%%%%%%%%%%%Figure 1%%%%%%%%%%%%%%%%%%%%%%%%%%%%%%%
\begin{figure}[!tbp]
{\includegraphics[height=1.5in,width=3.5in,angle=0]{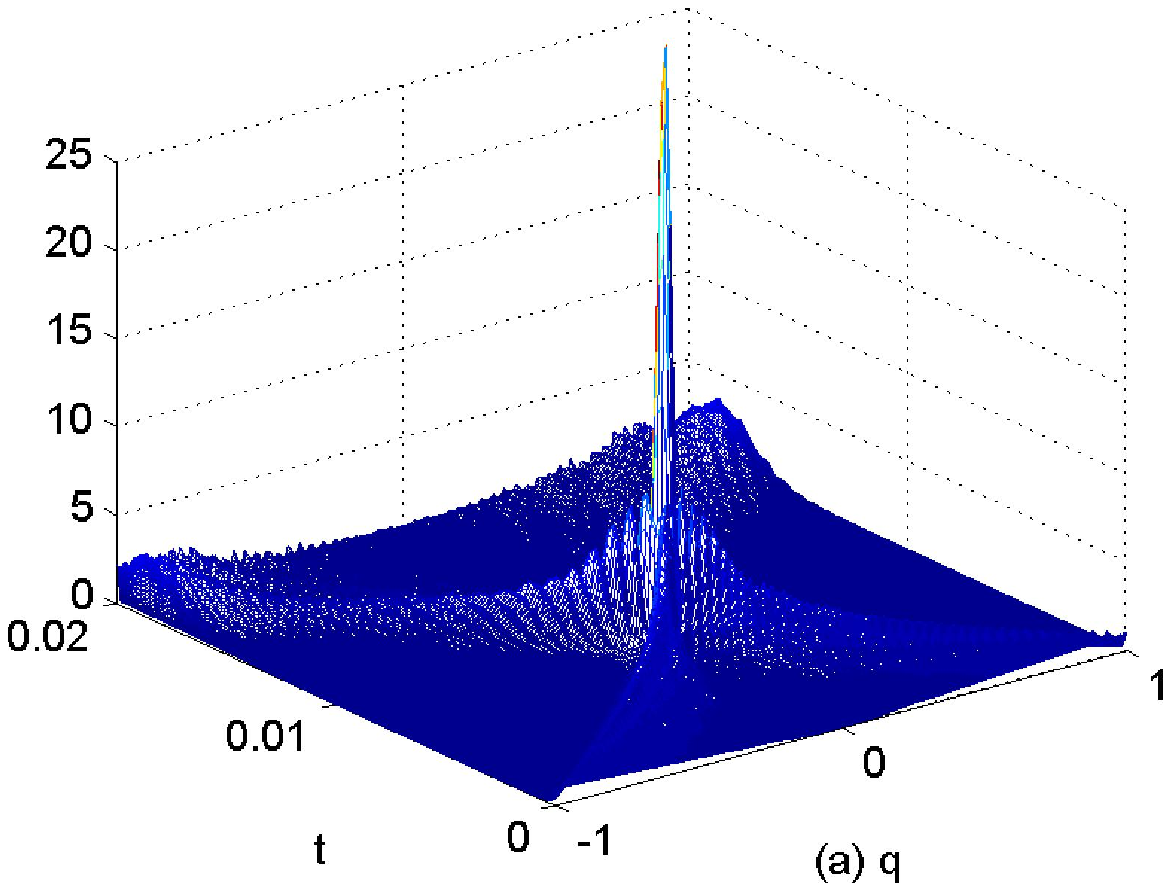}}
{\includegraphics[height=1.5in,width=3.5in,angle=0]{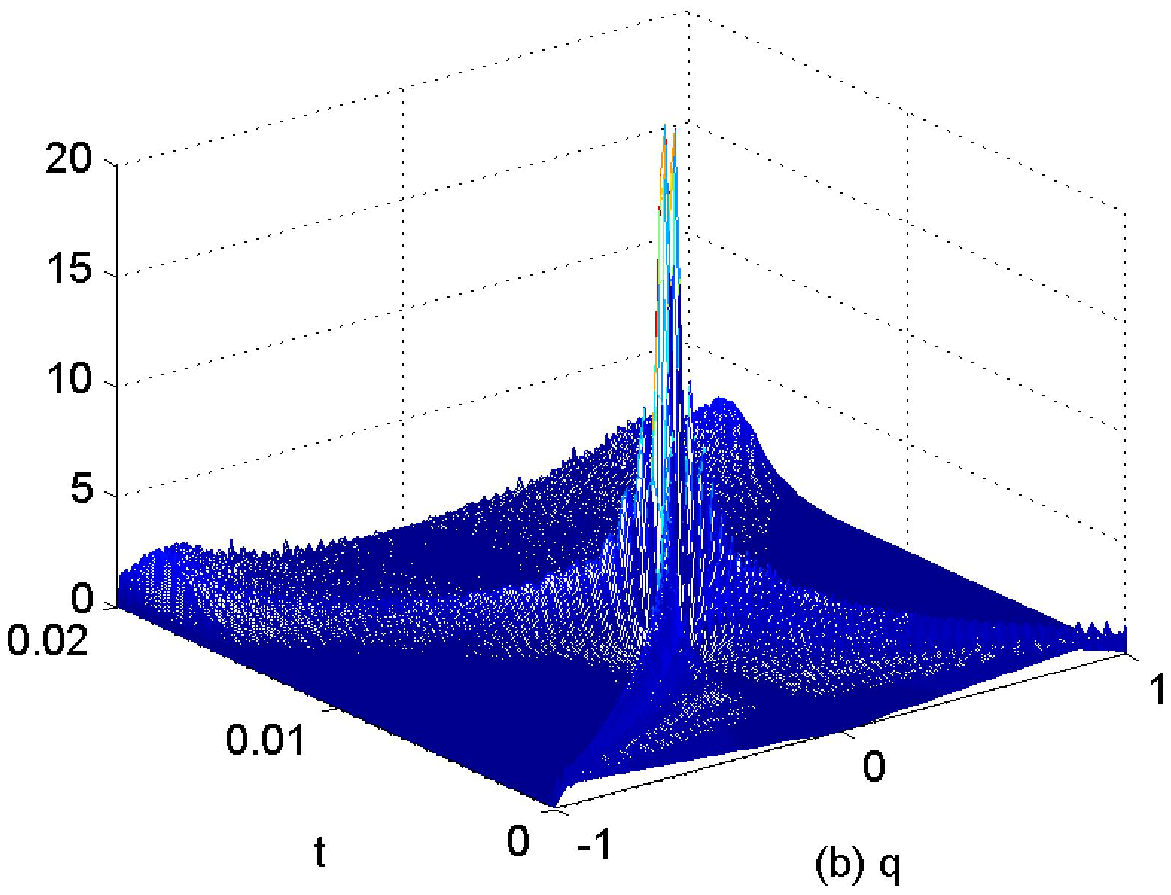}}
\caption{(Color online) The $n=20$ (a) and $n=21$ (b) evolved probability densities for $\gamma=0.01$, with $\hbar=l=m=1$. Both unitarily arrive at the origin at their respective eigenvalues, 0.0081 and 0.0079.} \label{fig:asym}
\end{figure}
%%%%%%%%%%%%%%%%%%%%%%%%%%%%%%%%%%%%%%%%%%%%%%%%%%%%%%%%%%%%%%%%

\begin{center}
\begin{table}[t]
\begin{center}
\begin{tabular}{c}\hline
Non-Periodic  CTOA-eigenfunctions $\gamma$=0.01\\
\begin{tabular}{|c|c|c|c|c|}\hline
$n$-th        & Eigenvalue       & Average          & Computed          &  Interpolated      \\
Eigen-        & ($\times 10^{-2}$) & Position         & minimum          &  Variance          \\
function      &                  & at the           &   variance               &  at the            \\
              &                  & Eigenvalue       &                  &  Eigenvalue        \\
              &                  & ($\times 10^{-4}$) & ($\times 10^{-2}$) &   ($\times 10^{-2}$) \\
\hline\hline
1             & 2887             &                  &                  &               \\
\hline
2             & 8.990            & -2.40            & 9.629            &  9.665           \\
\hline
3             & 7.161            & 1.19             & 6.367            &  7.208            \\
\hline
4             & 4.233            & -0.49            & 4.191            & 4.192             \\
\hline
5             & 3.758            & 0.35             & 3.361            & 3.586              \\
\hline
6             & 2.765            & -0.19            & 2.653            & 2.653              \\
\hline
7             & 2.551            & 0.16             & 2.264            & 2.366              \\
\hline
8             & 2.052            & -0.11            & 1.935            & 1.936              \\
\hline
9             & 1.931            & 0.09             & 1.697            & 1.758              \\
\hline
10            & 1.632            & -0.07            & 1.522            & 1.522              \\
\hline
\end{tabular}
\end{tabular}
\end{center}
\caption{Comparison of the calculated minimum variance with that of the linearly
interpolated variance at the eigenvalue for the corresponding eigenfunction for
the first ten eigenfunctions functions of the periodic CTOA-operator. The variances are
calculated with 601 Fourier coefficients at 0.0001 time steps.}
\end{table}
\end{center}

\subsection{The interpretation of the CTOA spectral properties}

The foregoing numerical results strongly endorse the minimal interpretation that a CTOA eigenfunction is a state that evolves to unitarily arrive
% or collapse\footnote{The collapse does not mean collapse to a singular support at the origin, but of obtaining the minimum variance (which is non-vanishing) of the position operator with respect to the eigenfunctions at their respective eigenvalues.} 
at the origin at its eigenvalue---that is, a state that evolves according to Schrodinger's equation such that the {\it event} of the position expectation value assumes the value zero, and the {\it event} of the position variance or uncertainty being minimum occur at the same instant of time equal to the corresponding eigenvalue. This provides the justification for the identification of the CTOA-operators as time of arrival operators.
%; and, for a non-parity eigenfunction, the expectation value of the position operator follows the classical trajectory at least from $t=0$ to just right after the eigenvalue. 
While our conclusion is based on a limited range of numerical results, these results are nevertheless compelling that we can take the minimal interpretation as an exact statement of the dynamical properties of the eigenfunctions of the confined time of arrival operators, until otherwise is proven.

But what is the importance of this realization? Recall that one of the surrounding issues in the past against the legitimacy of the quantum time of arrival problem within standard quantum mechanics is the absence of phase space trajectory for a quantum particle, so that the question of quantum time of arrival is ill-defined.
Our results here demonstrate that the QTOA-problem can be rephrased within the Hilbert space and the problem translates to finding states that unitarily arrive at a given point at a definite time. The QTOA-problem phrased in this way is  well-defined because quantum states have well-defined trajectories in the Hilbert space according to the Schrodinger equation.

But does this not trivialize the QTOA-problem because one can always construct by hand states that will evolve to have arbitrarily sharp width at the origin by exploiting the unitarity of quantum dynamics \cite{lamb0,lamb1}?
Constructing a single state with such properties is trivial, but the problem becomes non-trivial when we require the solution to comprise a complete set of such states, with the operator that can be constructed from this set being canonically conjugate with the Hamiltonian, and has an unambiguous classical limit which is the classical time of arrival \footnote{It is not clear at the moment if one can construct a complete set of orthonormal system from a set of CTOA-like wavefunctions constructed by exploiting the unitarity of quantum dynamics for closed quantum systems \cite{lamb0,lamb1}. This deserves further investigation.}. These requirements maybe necessary if we were to preserve the quantum-classical correspondence. 
The completeness requirement is inevitable if we require a theory of quantum first time of arrivals that is reflective of the fact that the entire phase space is accessible to a quantum particle via quantum tunneling.
Clarification of these issues will have to wait for the general theory of  confined quantum time of arrivals for arbitrary potentials, and the investigation of the limit as $l$ goes to infinity \cite{galapon5,fernando}.

%We point out though that while 

\section{Conclusion}\label{conclu}

In this paper we have given full account of the confined quantum time of arrival operators. While our results have answered some questions, they have raised some more questions and left others unanswered. Now we know self-adjointness and conjugacy of a time operator with a semi-bounded Hamiltonian can be achieved simultaneously; and a time operator need not be covariant and thus can be compact, with the eigenfunctions and eigenvalues tied with the dynamics of the system, acquiring interpretation independent from the quantum measurement postulate. 

However, these very realizations raise several fundamental questions. The dynamical interpretation of the eigenvalues of the CTOA-operators does not fit well with the fundamental quantum measurement postulate for observables represented by self-adjoint operators with discrete spectrum.
Does this imply a reconsideration of the fundamental quantum measurement postulate? In particular, does this call for a classification of observables according to their relationship or non relationship with the system dynamics? What sort of modification to the fundamental postulates to accommodate such classification if indeed necessary?
But then this leaves us with the question of the relationship between the spectral properties of a discrete time operator with the actual clicks of a detector.  Should these questions prove imperative their answers will have non-trivial repercussions at the foundational level. 

\section*{Appendix A: Derivation of the kernels}
The kernel $\left<q\right|T_{\gamma}\left|q'\right>$ for the non-periodic CTOA-operator is derived as follows. Using the property $\opr{q}\left|\left.q\right>\right.=q\ket{q}$, we have
\begin{eqnarray}
\left<q\right|T_{\gamma}\left|q'\right>&=&-\frac{\mu}{2}\left(q+q'\right)\left<q\right|\opr{p}^{-1}_{\gamma}\left|q'\right>\nonumber\\
&=&-\frac{\mu}{4\hbar}\left(q+q'\right)e^{\frac{i\gamma}{l}(q-q')}\!\!\sum_{k=-\infty}^{\infty}\frac{e^{\frac{i k\pi}{l}(q-q')}}{\gamma+k\pi}
\label{TOAkernel},
\end{eqnarray}
where the second line follows from introducing a resolution of unity provided by the complete eigenvectors of the momentum in the factor $\left<q\right|\opr{p}^{-1}_{\gamma}\left|q'\right>$. Now the sum can be rewritten in the form
\begin{eqnarray}
\sum_{k=-\infty}^{\infty}\frac{e^{\frac{i k\pi}{l}(q-q')}}{\gamma+k\pi}&=&\left\{\frac{1}{\gamma}+2\gamma\sum_{k=1}^{\infty}\frac{\cos\left[\frac{k\pi}{l}(q-q')\right]}{\gamma^{2}-k^{2}\pi^{2}}\right. \nonumber\\
& &\left.-2i\pi\sum_{k=1}^{\infty}\frac{k \sin\left[\frac{k\pi}{l}(q-q')\right]}{\gamma^2-k^2\pi^2}\right\}\label{kol}
\end{eqnarray}
The two infinite series can be straightforwardly evaluated by contour integration. It yields the explicit forms
\begin{eqnarray}
\sum_{k=1}^{\infty}\frac{\cos\left[\frac{k\pi}{l}(q-q')\right]}{\gamma^{2}-k^{2}\pi^{2}}&\!\!\!=\!\!\!&\frac{\cos\left[\gamma\left(1-\frac{|q-q'|}{l}\right)\right]}{2\gamma \sin\gamma}-\frac{1}{2\gamma^{2}},\label{cosinesum}\nonumber\\
\sum_{k=1}^{\infty}\frac{k \sin\left[\frac{k\pi}{l}(q-q')\right]}{\gamma^{2}-k^{2}\pi^{2}}\!\!\!&=&\!\!\!-\frac{\sin\left[\gamma\left(1-\frac{|q-q'|}{l}\right)\right]}{2\pi \sin\gamma}\mbox{sgn}(q-q')\label{sinesum}.\nonumber
\end{eqnarray}
Substituting these back into Equation-({\ref{kol}) and after some simplification, we have
\begin{eqnarray}
\sum_{k=-\infty}^{\infty}\frac{e^{\frac{i k\pi}{l}(q-q')}}{\gamma+k\pi}&=& \frac{\cos\left[\gamma\left(1-\frac{|q-q'|}{l}\right)\right]}{\sin\gamma}\nonumber\\
& & + i \frac{\sin\left[\gamma\left(1-\frac{|q-q'|}{l}\right)\right]}{\sin\gamma}\mbox{sgn}(q-q')\nonumber.
\end{eqnarray}
This can still be simplified,
\begin{equation}
\sum_{k=-\infty}^{\infty}\frac{e^{\frac{i k\pi}{l}(q-q')}}{\gamma+k\pi}
=\frac{1}{\sin\gamma}\left\{\begin{array}{ll} e^{-i\gamma\frac{(q-q')}{l}}\, e^{i\gamma}, & q>q' \\
                        \cos\gamma, & q=q'\\
                          e^{-i\gamma\frac{(q'-q)}{l}}\, e^{-i\gamma}, & q<q' \end{array}\right.
\end{equation}
Finally substituting this back into equation-(\ref{TOAkernel}) give us the kernel-(\ref{repre}), as long as we define $H(0)=1/2$.

The kernel $\left<q\right|T_{0}\left|q'\right>$ for the periodic CTOA-operator can be derived similarly. Again using the property $\opr{q}\ket{q}=q\ket{q}$ gives us
\begin{eqnarray}
\left<q\right|\opr{T}_{0}\left|q'\right>&&=-\frac{\mu}{2}(q+q')\left<q\right|\opr{P}^{-1}\left|q'\right>\nonumber\\
&&=-\frac{\mu}{4\hbar}\left(q+q'\right)\sum_{k=-\infty}^{\infty}\!\!\!'\frac{1}{k\pi}e^{\frac{ik\pi}{l}(q-q')}\nonumber\\
&&=-\frac{i\mu}{2\pi l\hbar}\left(q+q'\right)\sum_{k=1}^{\infty}\frac{1}{k}\sin\left[\frac{k\pi}{l}(q-q')\right]\nonumber.
\end{eqnarray}
Using the identity $\sum_{n=1}^{\infty}n^{-1}\sin x=\pi \mbox{sgn}(x)-x$ for all $-2\pi\leq x\leq 2\pi$, we finally arrive at the kernel-(\ref{periodic}).

\section*{Appendix B:\\Numerical Solution to the CTOA Eigenvalue Problem}

An independent numerical solution can be obtained for the CTOA-eigenvalue problem. Here we describe the Nystrom method of solving the Fredholm integral eigenvalue problem. Generally the integral operator eigenvalue problem is of the form
\begin{equation}\label{integral}
\int_{a}^{b}K(q,q')\,\varphi(q')\, dq'=\lambda \varphi(q),
\end{equation}
where $\varphi$ is an eigenfunction and $\lambda$ the corresponding eigenvalue, with $a<b$. The Nystrom method is based on some choice of an integration quadrature, $\int_a^b \psi(q)\, dq=\sum_{j=1}^N w_j \psi(q_j)$,
where the $w_j$'s are the weights of the quadrature rule and the N points $q_j$'s are the abscissa. 

Using the quadrature rule, equation-(\ref{integral}) reduces to $\sum_{j=1}^N K(q,q_j) w_j \varphi(q_j) = \lambda \varphi(q)$, and evaluating this expression at the absiccas further reduces to $\sum_{j=1}^N K(q_i,q_j) w_j \varphi(q_j) = \lambda \varphi(q_i)$,
for $i=1, 2, \dots$. Numerically the Fredholdm eigenvalue problem then reduces to the matrix eigenvalue problem
\begin{equation}
\tilde{K}\cdot \tilde{\varphi}=\lambda\tilde{\varphi}
\end{equation}
where $\tilde{K}_{i,j}=K(q_i,q_j) w_j$ and $\tilde{\varphi}_i=\varphi(q_i)$. The $N$ eigenvalues and corresponding $N$ eigenfunctions are the approximations to the first $N$-th largest eigenvalues and the corresponding eigenfunctions of the Fredholm integral eigenvalue problem.

Appropriate to our problem at hand is Gauss-Legendre integration quadrature in the interval $[-1,1]$. This is possible for any $l$ because we can always rescale the interval $[-l,l]$ to the interval $[-1,1]$. The $N$ abscissas are the $N$ roots of the Legendre polynomial
\begin{equation}
P_N(x)=\frac{1}{2^N}\sum_{k=0}^{\left\lfloor N \right\rfloor}(-1)^k \left(\stackrel{N}{k}\right) \left(\stackrel{2N-2k}{N}\right) x^{l-2k}
\end{equation}
in the interval $[-1,1]$. The weights are given by
\begin{equation}
w_j=\frac{\int_{-1}^1 P_n (x)^2 dx}{P_{N-1}(x_j) P_N'(x_j)}
\end{equation}
where $P_N'(x_j)$ is the derivative of the Legendre polynomial at its zero $x_j$.

Table-IV shows the comparison of the exact and the numerical results for the eigenvalues of the CTOA-operators for the indicated values of $\gamma$. The eigenvalues have been computed using Nystrom method and employing Gauss-Legendre quadrature with $10000$ integration points. There is an excellent agreement between the exact and the numerical values for the CTOA-operators with definite parity, i.e. for $\gamma = 0,\frac{\pi}{2}$; and a fair agreement for other values of $\gamma$.

%Work out here the table
\begin{table}[h]\label{comp}
	\begin{tabular}{||c|c|c|c|c|c|c||} \hline
	n  & \multicolumn{2}{|c|}{$\gamma=0$} & \multicolumn{2}{|c|}{$\gamma=\frac{\pi}{8}$}& \multicolumn{2}{|c|}{$\gamma=\frac{\pi}{4}$}\\\cline{2-7}
	    & exact & numeric  & exact & numeric  & exact & numeric\\\hline
1&0.124608&0.124608&0.73965&0.73953&0.37834&0.37829\\\hline
2&0.111438&0.111438&0.09459&0.09468&0.10510&0.10515\\\hline
3&0.050180&0.050180&0.06887&0.06878&0.06415&0.06411\\\hline
4&0.048799&0.048799&0.04332&0.04341&0.04539&0.04543\\\hline
5&0.031585&0.031585&0.03683&0.03674&0.03545&0.03541\\\hline
6&0.030274&0.030274&0.02806&0.02815&0.02892&0.02896\\\hline
7&0.022849&0.022849&0.02516&0.02507&0.02451&0.02447\\\hline  
\end{tabular}
	\caption{Comparison of exact values of the first 7 eigenvalues of the CTOA operator for $\gamma=0$, $\gamma=\frac{\pi}{8}$ and $\gamma=\frac{\pi}{4}$ with the eigenvalues of the CTOA operator for $\gamma=0$, $\gamma=\frac{\pi}{8}$ and $\gamma=\frac{\pi}{4}$ computed using the Nystrom method.}
\end{table}

\section*{Acknowledgment}
This work has been supported by the National Research Council of the Philippines  through grant number I-81-NRCP, and partially supported by the ``Ministerio de Ciencia y Technologia'' and FEDER (Grant BFM2003-01003). EAG is supported by the University of the Philippines through the U.P. Creative and Research Scholarship Program. This paper has benefited from discussions with I. Egusquiza, and J. G. Muga.

\end{document}